\documentclass[aps,prd,nofootinbib,floatfix]{revtex4}
\usepackage{epsfig}
\usepackage{amsmath,amsfonts}
\usepackage{subfigure}

\def\Tr{\mathrm{Tr}}

\def\tmchpt{tm$\chi$PT}
\def\LQ{\Lambda_{\rm QCD}}

\def\spose#1{\hbox to 0pt{#1\hss}}
\def\ltapprox{\mathrel{\spose{\lower 3pt\hbox{$\mathchar"218$}}
 \raise 2.0pt\hbox{$\mathchar"13C$}}}
\def\gtapprox{\mathrel{\spose{\lower 3pt\hbox{$\mathchar"218$}}
 \raise 2.0pt\hbox{$\mathchar"13E$}}}
\def\inapprox{\mathrel{\spose{\lower 3pt\hbox{$\mathchar"218$}}
 \raise 2.0pt\hbox{$\mathchar"232$}}}

\begin{document}
%\preprint{UW-PT 05-??} %%%not used by the UW any more

\title{Observations on discretization errors in twisted-mass lattice QCD}

\author{Stephen~R.~Sharpe}
\email[]{sharpe@phys.washington.edu}
\affiliation{Physics Department, University of Washington,
Seattle, WA 98195-1560, USA}

\date{\today}

\begin{abstract}
I make a number of observations concerning discretization 
errors in twisted-mass lattice QCD
that can be deduced by applying chiral perturbation theory
including lattice artifacts. (1) The line along
which the PCAC quark mass vanishes in the  untwisted mass-twisted mass
plane makes an angle to the twisted mass axis which is a direct
measure of $O(a)$ terms in the chiral Lagrangian, and is found 
numerically to be large; (2) Numerical
results for pionic quantities in the mass plane show the qualitative
properties predicted by chiral perturbation theory, in
particular an asymmetry in slopes between positive and
negative untwisted quark masses;
(3) By extending the description of
the ``Aoki regime'' (where $m_q\sim a^2 \Lambda_{\rm QCD}^3$)
to next-to-leading order in chiral perturbation theory I show how
the phase transition lines and lines of maximal twist (using different
definitions) extend into this region, and give predictions
for the functional form of pionic quantities; (4) I argue that
the recent claim that lattice artifacts at maximal twist have 
apparent infrared
singularities in the chiral limit results from expanding
about the incorrect vacuum state. Shifting
to the correct vacuum (as can be done using chiral
perturbation theory) the apparent singularities are summed into non-singular,
and furthermore predicted, forms. I further argue
that there is no breakdown in the Symanzik expansion in powers of lattice
spacing,
% (in the sense that the contributions of all orders
% beyond $O(a^2)$ lead to corrections suppressed by explict powers of $a\LQ$),
and no barrier to simulating at maximal twist in the Aoki regime.

\end{abstract}

\maketitle

%-------------------------------------------------------------- SLIDE -
\section{\label{sec:outline} Introduction}
Twisted mass fermions~\cite{tm} at maximal twist~\cite{FR1,FR2,FR3} are 
 presently under intensive study
both with simulations and theoretical methods.\footnote{% 
For recent reviews see Refs.~\cite{FrezzottiLat04,ShindlerLat05}.
}
The standard application involves two degenerate quarks, and this is the
theory considered here.\footnote{% 
Extension to non-degenerate quarks has been explained in
Ref.~\cite{FRnondegen}.}
The long distance physics (vacuum and pionic properties) of twisted
mass lattice QCD with two degenerate flavors (tmLQCD) can be studied
using the methods of effective field theory, allowing
a systematic dual expansion in the quark mass, $m_q$, and the
lattice spacing, $a$, in which non-perturbative effects are
parameterized by {\em a priori} unknown 
low energy constants (LECs)~\cite{ShSi,Noam,Oliver}.
The resulting theory is called twisted-mass chiral perturbation theory
(\tmchpt)~\cite{MS}.
Its consequences in the vacuum and pionic sector have been
worked out to next-to-leading order (NLO)
in the regime with $m_q \gtapprox a\Lambda^2$ (the ``generic small
mass'' or GSM regime\footnote{%
Note that the GSM regime is here defined to include the region
in which $m_q \gg a\Lambda^2$, since the NLO expressions remain
valid in this region, although the terms proportional to $a^2$ will
become of NNLO. Of course, for chiral perturbation theory to apply
one must always have that $m_q \ll \Lambda$. In practice, for present
lattice spacings, this latter condition implies that $m_q$
does not exceed $a\Lambda^2$.}
)~\cite{AB04,ShWu2}, and at leading order (LO)
in the regime with $m_q \sim a^2\Lambda^3$ (the ``Aoki'' 
regime)~\cite{Mun04,Scor04,AB04,ShWu1,ShWu2}. 
Here $\Lambda$ is a scale of the order of the QCD scale, $\LQ$.

The present paper is a collection of observations that follow from
\tmchpt\ concerning discretization errors in tmLQCD.
On the practical side, I point out in sec.~\ref{sec:maxtwist}
some new applications of results contained in Ref.~\cite{ShWu2}. 
In particular, I describe how the
observed difference between two methods of determining maximal twist
that are being used in simulations can be understood using \tmchpt. 
This difference provides
a direct measure of the size of discretization errors, or equivalently
a measure of the LECs related to such errors. 
I also point out the relationship of these methods to a third 
(theoretically interesting but apparently less practical) 
method proposed in Ref.~\cite{ShWu2}.

In sec.~\ref{sec:fits} I note that the dependence of physical quantities
on the untwisted quark mass at fixed twisted quark mass 
gives another set of measures of the size
of discretization errors, some of which are related by \tmchpt\ at NLO.
These results are contained in Ref.~\cite{ShWu2}, but have not been compared
to the available numerical data, and I think it is useful to
provide illustrative plots of the expected types of dependence.

When making these plots the question arises as to how to extend them into
the Aoki regime, which is the regime in which there are phase transitions
caused by competition between quark mass effects and those due to
discretization errors.
Previous studies, in Refs.~\cite{Mun04,Scor04,AB04,ShWu1,ShWu2},
 worked to LO in this regime,
and here I extend these results to NLO. This extension, described
in sec.~\ref{sec:aoki}, turns out to require only one non-trivial
additional low-energy coefficient.
The results allow one to see how the phase transition lines,
which are linear at LO, can become curved at NLO, and similarly
how the lines of maximal twist using different definitions behave.
These results may be amenable to numerical investigation,
but, in any case, are of theoretical interest in understanding
the relationship between different definitions of twist angle and
critical mass.
I also point out that, while a classical analysis is sufficient when
working to NLO in the Aoki regime, extension to next-to-next-to-leading order
(NNLO) requires the inclusion of loop effects.

My final topic, discussed in sec.~\ref{sec:IR},
is the conclusions of Ref.~\cite{FMPR}.
It is argued in \cite{FMPR} that, if one makes an estimate
of the critical mass that has an error of size $a\Lambda^2$, then,
when working at maximal twist, physical
correlation functions contain apparently
infrared (IR) divergent discretization errors
proportional to $(a/m_{\pi^0}^2)^{2k}$, with $k\ge 1$ an integer.
Such errors, while consistent with automatic $O(a)$ improvement,
 would restrict one to working with $m_q > a\Lambda^2$,
and suggest a breakdown of the Symanzik expansion~\cite{Symanzik}
 for smaller quark masses.
It is then argued that, either by using an ``optimal'' choice of critical mass,
or using a non-perturbatively improved quark action, these IR divergent
errors at maximal twist
can be reduced in size to $a^2(a^2/m_{\pi^0}^2)^k$. In this way
the bound on the quark mass is reduced to $m_q > a^2\Lambda^3$,
i.e. one can work in the GSM regime but not in the Aoki regime.

My observation is that the IR divergences are
an indication of expanding around the wrong vacuum in
the presence of large perturbation. Using \tmchpt\ one can sum these
corrections to all orders and obtain a finite and smooth prediction
for the dependence of physical quantities on the underlying untwisted
and twisted quark masses. Indeed, this is precisely what was done previously
in Refs.~\cite{AB04,ShWu2}. 
I argue further that there is no breakdown of the Symanzik expansion,
suitably interpreted,
and no barrier to working at maximal twist in the Aoki regime.

These considerations do not, however, invalidate the conclusion of
Ref.~\cite{FMPR} that one should use an appropriate
``optimal'' choice of critical mass to reduce the size
of discretization errors. This was proposed previously
using \tmchpt\ in Refs.~ \cite{AB04,ShWu2}, and the criteria for
determining the optimal choice agree. My observation here is that,
since \tmchpt\ is built upon the Symanzik expansion used by Ref.~\cite{FMPR},
the result concerning optimal critical masses had to agree.
In fact, \tmchpt\ goes beyond the Symanzik expansion by adding
non-perturbative information concerning spontaneous chiral symmetry breaking.
The only additional assumption
that one has to make is that the effective field theory framework
is valid~\cite{Weinberg,Leutwyler}.

\bigskip
It is useful to keep in mind during the subsequent discussions the
physical values of quark masses that correspond to the different regimes.
If one takes $a^{-1}=2\;$GeV (a relatively small lattice spacing for
present dynamical simulations) and $\Lambda\approx\LQ\approx 300\;$MeV,
then $a\Lambda^2$ (the size of quark masses in the GSM regime)
and $a^2\Lambda^3$ (the size in the Aoki regime)
are about $45$ and $7\;$MeV respectively.
Since we are interested in calculating with light quark masses varying from 
$m_{\rm strange} \approx 80-100\;$MeV down toward $m_{\rm light}\approx 3-4\;$MeV,
 most simulations will be done in the GSM regime,
with a significant tail entering the Aoki regime.
Thus it is important to understand both regimes in detail.

\medskip

In this paper I work in the ``twisted basis'' in which the
tmLQCD action is
\begin{equation}
\label{E:action}
S^L_F = \; \sum_{x} \bar{\psi_l}(x)
\Big[\frac{1}{2} \sum_{\mu} \gamma_\mu (\nabla^\star_\mu + \nabla_\mu)
- \frac{r}{2} \sum_{\mu} \nabla^\star_\mu \nabla_\mu 
+ m_0 + i \gamma_5 \tau_3 \mu_0 \Big] \psi_l(x),
\,.
\end{equation}
Here $\psi_l$ is the dimensionless bare lattice field,
$\mu_0$ is the bare twisted mass and $m_0$ the bare untwisted mass.
The coefficient of the Wilson term, $r$, is typically set to unity in simulations;
the considerations that follow hold, however, for any $O(1)$ choice of this
coefficient. Maximal twist is obtained by setting $m_0$ to an estimate
of its critical value $m_c$ (which, as will be seen, can depend on $\mu_0$).
 The determination of $m_c$ is a key
issue in tmLQCD, and the topic of much of the discussion which follows.
I only note here that this is a non-trivial issue as no symmetry is
restored when $m_0=m_c$. By contrast, setting $\mu_0=0$ restores parity
and flavor symmetries, and physical quantities are symmetrical 
under $\mu_0\to-\mu_0$.

I use the twisted basis, rather than the more continuum-like 
``physical basis'', used for example in Ref.~\cite{FMPR},
since the twisted basis is that usually used in simulations,
allowing a more direct connection to the numerical data. 
It is also the basis used in the \tmchpt\ calculations of
Ref.~\cite{ShWu2}, which I use extensively.

The physical twisted and
untwisted masses are $\mu = Z_\mu \mu_0/a$ and $m = Z_m (m_0/a-m_c)$, respectively,
with $Z_\mu=1/Z_P$ and $Z_m=1/Z_S$ renormalization constants.\footnote{%
Note that I define $m_c$ to be dimensionful.}
The physical quark mass
is then $m_q =\sqrt{\mu^2 + m^2}$, and to work at
maximal twist requires setting $m=0$ (i.e. setting $m_0=a m_c$).
The accuracy with which $m_c$ must be determined in order to 
maintain automatic $O(a)$ improvement depends on how close to
the chiral limit one is working. If $m_q \gg a\Lambda^2$, then discretization 
errors are small, and the error in $m_c$ can be of $O(a)$.\footnote{%
Here, and in the following, I often use a notation in which appropriate
factors of $\Lambda$ are implicit.}
This possibility is part of the topic of sec.~\ref{sec:IR}.
In the GSM regime, one must incorporate $O(a)$ discretization effects
into $m_c$~\cite{AB04,ShWu2}, and thus reduce the error to $O(a^2)$.
To work in the Aoki regime and maintain automatic $O(a)$ improvement
requires the uncertainty in $m_c$ to be further reduced, 
to $O(a^3)$~\cite{AB04,ShWu2}. 
Following Ref.~\cite{FMPR}, I refer to choices of critical mass
which are accurate to $O(a^3)$ as ``optimal'', and
denote them by $m'_c$ (and the resulting untwisted quark mass by $m'$). 
I discuss these choices in secs.~\ref{sec:maxtwist} and \ref{sec:fits}.
One noteworthy choice of $m_c$, which, as stressed
in Ref.~\cite{AB04}, is  not optimal,
is given by the position where the pion mass vanishes 
(an end-point of the Aoki phase, if present).
This differs from optimal choices by $O(a^2)$.
Finally, if one works at NLO in the Aoki regime,
it is natural, as shown in sec.~\ref{sec:aoki}, to incorporate
an $O(a^3)$ shift into $m_c$. I denote the resulting
critical mass and untwisted physical quark mass by $m''_c$ and $m''$,
respectively. The uncertainty in $m''_c$ turns out to be of $O(a^5)$.
The fact that only odd powers of $a$ appear has been noted also
in Ref.~\cite{FMPR}.

So as not to clutter the notation with too many ``primes'', I use the
same (unprimed) symbol for $m_q$ in each regime, although it is defined
using the untwisted quark mass appropriate to the given regime.
Thus, for example, in the Aoki regime,
$m_q =\sqrt{\mu^2 + m''^2}$. 

The variables $m_0/a$ and $\mu_0/a$
(or $m$ and $\mu$, or $m'$ and $\mu$, etc.) map out what I
will refer to as the ``mass plane''. The origin in this plane
is defined to be where $\mu_0=0$ and $m_0/a$ is set
to the critical value appropriate to the regime
of interest. With the definitions given above,
$m_q$ is then the distance from the origin. Once $m_c$ is estimated,
one possible definition of the twist angle (that which relates most
directly to that used in the continuum) is the angle relative to the 
positive untwisted mass axis. This angle depends on the choice of
critical mass, but I refer to it as $\omega_0$ in all regimes. 
It is most useful in the GSM regime, where it is defined by
\begin{equation}
\omega_0\equiv \tan^{-1}(\mu/m')\,.
\label{eq:omega0def}
\end{equation}
In the Aoki regime, at NLO, one replaces $m'$ with $m''$.

Finally, I collect the notation I use
for low energy coefficients (LECs) in \tmchpt,
largely taken over from Ref.~\cite{ShWu2}.
At leading order, there are two
continuum LECs: $f$, the decay constant
in the chiral limit normalized to $f_\pi=132\;$MeV, and $B_0$,
defined so that $m_\pi^2 = 2 B_0 m_q$. For quark masses renormalized
at $2\;$GeV in the $\overline{{\rm MS}}$ scheme, $B_0\approx 2.5\;$GeV.
It is then convenient to
introduce $\hat\mu= 2 B_0 \mu$ and $\hat{m} = 2 B_0 m$ 
and $M = \sqrt{\hat\mu^2 + \hat{m}^2}$ so that, at LO, $m_\pi^2=M$.
At NLO in the continuum, there are the standard Gasser-Leutwyler
coefficients, the $L_i$~\cite{GLoneloop}.
Discretization errors introduce other LECs. At LO in the GSM regime, 
there is a single constant $W_0$, 
which determines the $O(a)$ shift in $m_c$~\cite{ShSi,Noam}. In particular, 
defining $\hat{a}= 2 W_0 a$, 
then the shifted untwisted mass is
$\hat{m}'\equiv 2 B_0 m' =\hat{m}+\hat{a}$, and $m_\pi^2=M'\equiv\sqrt{\hat\mu^2+(\hat{m}'){}^2}$.
Discretization errors of NLO in the GSM regime are determined by
the dimensionless constants $W$, $W_{10}$, $\widetilde W$, 
and $W'$~\cite{Noam,Oliver,ShWu2}.
It is convenient to introduce the following quantities:
\begin{equation}
\delta_W\equiv \frac{16 \hat{a} (W+W_{10}/4)}{f^2}\,,\ \
\delta_{\widetilde W} \equiv \frac{16 \hat{a} (\widetilde{W}+W_{10}/2)}{f^2}
\,,\ \
w' \equiv \frac{16 \hat{a}^2 W'}{f^2}\,.
\label{eq:deltaWdef}
\end{equation}
$\delta_W$ and $\delta_{\widetilde{W}}$ give dimensionless measures of
the size of discretization errors in physical quantities, while the mass-squared
$w'$ appears additively in the result for the pion mass-squared.
Note $\delta_W$ and $\delta_{\widetilde{W}}$ are linear in $a$,
while $w'$ is quadratic.
Further LECs enter at NLO in the Aoki regime, and will be defined
in sec.~\ref{sec:aoki}.

\section{\label{sec:maxtwist} Comparing definitions of maximal twist}

As noted in the introduction, one must use a non-perturbative method
of sufficient accuracy in order to determine when one is at maximum twist.
This section concerns the relation between two methods being used
in present simulations [which I refer to as methods (i) and (ii)]
and their relationship to a third method [method (iii)]
proposed in Ref.~\cite{ShWu2}.
Throughout this section I assume that the quark masses are in the GSM
regime.

\begin{figure}[t]
\begin{center}
\epsfxsize=4truein %\hsize
\epsfbox{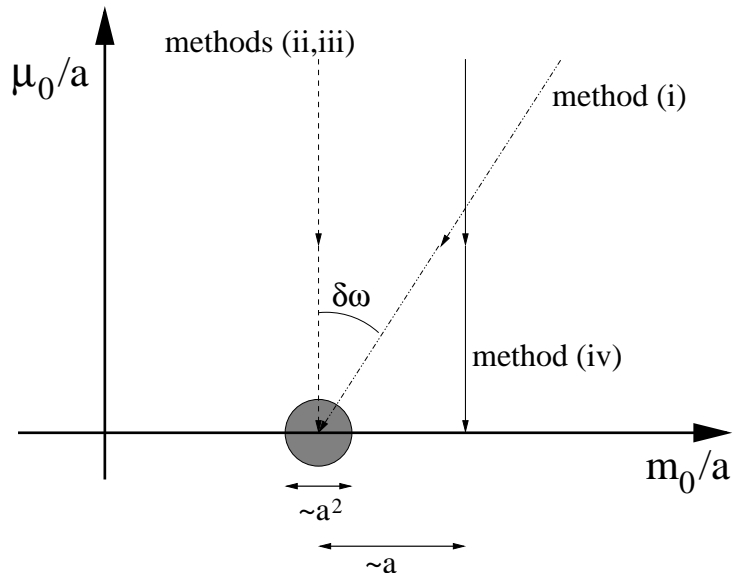}
\end{center}
\caption{Illustration of different methods for working at maximal twist.
See text for description [method (iv) is discussed
in sec.~\protect\ref{sec:IR}]. The arrows represent the direction one moves
to approach the chiral limit (which occurs at $\mu_0=0$). The plot 
represents the GSM regime in which $m_q\sim a$. The Aoki regime,
in which $m_q\sim a^2$ is represented by the shaded region.
It is discussed in sec.~\protect\ref{sec:aoki}. The angle $\delta\omega$
is defined to be positive for the situation shown in the Figure.}
\label{fig:maxtwist}
\end{figure}

My observations are summarized in Fig.~\ref{fig:maxtwist}.
The line in the mass-plane defined by
method (i) (in which the PCAC mass is set to zero) makes an angle $\delta\omega$
to the twisted mass direction [and thus, by construction,
 to the line defined by method (ii)]. This angle is proportional to the lattice
spacing, and to one combination of the LECs describing
discretization errors. It can be determined from present simulations, 
and the results show that this discretization error is not as small 
as one might hope, since it is
determined by a scale $\Lambda \approx 0.7\;$ GeV rather 
than by $\Lambda\approx \Lambda_{\rm QCD}$.
I also observe, following Ref.~\cite{ShWu2}, that method (iii), 
previously argued to be difficult to implement in practice, 
is actually equivalent to method (ii) in the GSM regime
(and thus has already been used). 

\bigskip

Both method (i) and (ii) are based on the continuum result that,
at maximal twist, charged axial currents in the twisted basis
correspond to physical vector currents, and thus should have
vanishing coupling to physical charged pions. It is useful to
introduce a definition of twist angle incorporating this result:
\begin{equation}
\tan\omega_A = \frac{\langle 0 | V^2_\mu| \pi^1\rangle}
                    {\langle 0 | A^1_\mu| \pi^1\rangle}
\,,\label{eq:omegaAdef}
\end{equation}
where the currents are in the twisted basis, and the superscripts
are adjoint $SU(2)$ indices. This angle is denoted $\omega$ in
Ref.~\cite{ShWu2}, but I add the subscript ``A'' here for clarity.
Maximal twist can then be defined by $\omega_A=\pm\pi/2$,
corresponding to~\cite{Farchioni04B}
\begin{equation}
\langle 0 | A^{1}_\mu | \pi^{1} \rangle = 0 \,,
\label{eq:PCACcond}
\end{equation}
Note that the charged pions can be unambiguously created
by the local pseudoscalar density (which is invariant under
the axial transformation that rotates between twisted and
physical bases). The condition (\ref{eq:PCACcond}) is equivalent to
setting the PCAC mass (defined in sec.~\ref{sec:fits} below) to zero.

The two methods apply the condition (\ref{eq:PCACcond}) in different ways.
In method (i) the condition is enforced for each
choice of $\mu_0$ by varying $m_0$~\cite{ShWu2,Lewis}. 
This leads to a curve in the
mass plane which one follows as $\mu_0$ is reduced.
To good approximation this curve is found to be a straight line 
for the masses used in present simulations~\cite{Lewis,Jansen05}.
This method is sometimes called the parity-conservation definition,
but I do not use this name as parity is not restored along the line
it defines at non-zero lattice spacing [and the name could equally well be 
applied to method (iii) discussed below].
Method (ii) defines the
critical mass by extrapolating the curve along
which the condition (\ref{eq:PCACcond}) holds to $\mu_0=0$~\cite{Jansen05}.
The resulting simulations are then done holding $m_0$ fixed
at this value, with $\mu_0$ varying.
This method is also that proposed in Ref.~\cite{FMPR}.
Figure~\ref{fig:maxtwist} illustrates the difference between 
the two methods.\footnote{%
Both methods (i) and (ii) [as well as method (iii) discussed below]
are optimal in the sense discussed in the
introduction---they lead to automatic $O(a)$ improvement even in the
Aoki regime. This property is not important, however, in this section.
}

The curve defined in method (i) is displayed in Fig.~7 of Ref.~\cite{Lewis}
and Fig.~1 of Ref.~\cite{Jansen05}. (In both figures the $m_0$ and $\mu_0$ axes
are interchanged compared to Fig.~\ref{fig:maxtwist}, and in the second figure the
vertical axis is $1/\kappa$ rather than $m_0=1/(2\kappa)-4$.)
These are results from quenched simulations, but I will not rely on the
details of these results, instead assuming only that the following features
are general and continue to hold in unquenched simulations:
(1) the curves are almost linear for small $\mu$
(roughly for $\mu_0/a \ltapprox 30\;$MeV); and (2) they approach the $\mu_0=0$ axis
at an angle that differs from $\pi/2$. Indeed this difference (denoted by $\delta \omega$
in Fig.~\ref{fig:maxtwist}) is significant: approximately $0.8$ radians for $\beta=5.7$~\cite{Jansen05},
$0.5$ for $\beta=5.85$ and $0.35$ for $\beta=6$~\cite{Lewis}.
%%
%% These numbers are eyeballed from the figures.
%%

The first observation I want to make is that the result $\delta\omega\ne 0$
is a prediction of \tmchpt, and that the value of $\delta\omega$ is a direct measure
of $O(a)$ discretization errors. This follows from results 
contained in Ref.~\cite{ShWu2}. In particular, method (i) is the same as using the
canonical definition of maximal twist suggested in Ref.~\cite{ShWu2}. It follows
from setting $\omega_A=\pi/2$ in eq.~(58) of that work 
(where, recall, $\omega_A$ is denoted $\omega$) that
the line of maximal twist in method (i) lies at an angle
\begin{equation}
\omega_0 = \pi/2 - \delta\omega\,,\qquad
\delta\omega = - (Z_m/Z_\mu)\delta_W + O(a^2)
\,,
\label{eq:deltaomega}
\end{equation}
relative to the origin where $\mu_0=0$ and $m_0/a=m'_c$
[i.e. $\omega_0$ is defined as in eq.~(\ref{eq:omega0def})].
Here $\delta_W$ is defined in eq.~(\ref{eq:deltaWdef}) above,
and the factor of $Z_m/Z_\mu=Z_P/Z_S$ arises because the predictions of
Ref.~\cite{ShWu2} are given in terms of physical masses, rather than the
bare masses used in Fig.~\ref{fig:maxtwist}.
An implication of the result (\ref{eq:deltaomega}) is that the
line defined in method (i) extrapolates to a critical mass which
includes the $O(a)$ shift, confirming that it is an optimal choice.
It then follows that method (ii) corresponds to setting $\omega_0=\pi/2$.

The most noteworthy feature of eq.~(\ref{eq:deltaomega}) is that the
effect is linear in $a$. The angle $\delta\omega$
 is an example of a quantity which is
not automatically improved at maximal twist (as discussed more extensively
in Ref.~\cite{ShWu2}). The result (\ref{eq:deltaomega})
also incorporates the result that using different
lattice axial currents in the criterion (\ref{eq:PCACcond})
leads to different results for $\delta\omega$. For example,
using an improved current with $c_A\ne 0$ instead of an unimproved
current can be seen to
lead to a change in $\delta\omega$ proportional to $a c_A$.\footnote{%
I thank Andrea Shindler for alerting me to this dependence.}
This effect enters through the dependence of $\delta_W$ on
the LEC $W_{10}$, for $W_{10}$ itself depends on the choice of current.
Note that if one uses non-perturbatively improved Wilson fermions
(which sets $W=0$ as discussed in Ref.~\cite{ShWu2}) and
a non-perturbatively improved axial current (which sets $W_{10}=0$),
then $\delta\omega$ is predicted to vanish at $O(a)$.
In this case the difference between methods (i) and (ii) appears
at next-to-next-to-leading order (NNLO) in \tmchpt, arising, for example,
from a contribution proportional to $a \mu^2 \bar\psi\psi$ in the
Symanzik effective action. This latter term gives the critical mass a quadratic
dependence on $\mu_0^2$. The resulting small curvature 
can in fact be seen in
the quenched results discussed above, particularly in the results at
the largest lattice spacing~\cite{Jansen05}.

It should be stressed that a non-zero value for $\delta\omega$ does not
contradict automatic improvement as long as it is of $O(a)$. 
Contributions to physical quantities
linear in $a$ are necessarily also proportional to $\delta\omega$, 
so the overall discretization error
remains of $O(a^2)$. Indeed, as stressed in Refs.~\cite{FR1,ShWu2}, there is
an intrinsic, irreducible ambiguity in the twist angle: different criteria,
equally good in the continuum limit, lead to results differing at $O(a)$
at non-vanishing lattice spacing.
In fact, one can define maximal twist as lying along
any line in the mass plane whose angle satisfies $\omega_0=\pi/2 + O(a)$, where
the $O(a)$ offset is arbitrary. Note that for all such lines the
quark mass is $m_q = \mu [1 + O(a^2)]$, and thus, to NLO accuracy, one
can use $m_q=\mu$.

The above-mentioned curvature in the maximal twist line of method (i) 
allows one to estimate the size of the ambiguity in the resulting critical
mass. The NNLO curvature means that, when extrapolating
to $\mu=0$ from $\mu\sim O(a)$, the angle $\delta \omega$ has
an error of $O(a^2)$. This leads to an error in the
intercept $m'_c$ of size $O(a^3)$, consistent with the size predicted
by \tmchpt\ and noted in the introduction.

For practical applications it is important to study
$\delta\omega$ as an indicator of the size of discretization
errors. One expects $|\delta\omega|= a \Lambda$, with $\Lambda$ a scale
lying somewhere in the range from $\Lambda_{\rm QCD}\approx 300\;$MeV
to $\Lambda_\chi = 2\sqrt{2} f\approx 1.2\;$GeV.
For the $\beta=6$ results of Ref.~\cite{Lewis}
($\delta\omega\approx 0.35$ at $a^{-1}\approx 2\;$GeV)
one finds $\Lambda\approx 0.7\;$GeV (assuming $Z_\mu/Z_m\approx 1$).
The results at the other lattice spacings are roughly consistent
with this scale, or, to put it another way, are roughly consistent
with the expected linear dependence on $a$. One should not expect
more than a rough consistency because there is additional implicit
logarithmic dependence on $a$, entering, for example, through
the factor $Z_\mu/Z_m$, as well as NNLO terms of $O(a^2)$.

Given the result for $\delta\omega$, one can very roughly estimate
the size expected for the $O(a^2)$
errors in physical quantities---they should be approximately be of
relative size $(\delta\omega)^2$.
This is approximate because different LECs and numerical factors enter for
each quantity. Nevertheless, if one uses it as a guide, one expects
$O(a^2)$ errors of relative size $\sim 10\%$ for $a^{-1}\approx 2\;$GeV.\footnote{%
This applies for quantities that do not vanish in the chiral limit,
such as $f_\pi$ and $m_N$. For the pion mass-squared, and in particular
for the minimum value of the charged pion mass-squared in the case
of the first-order transition scenario, or
the splitting between charged and neutral pion mass-squareds, the
$O(a^2)$ effects are additive, and of size $a^2\Lambda^4$.}
It should be borne in mind, however, that changes to the gluon and fermion actions,
as well as changing from quenched to unquenched simulations, will change the
size of the LECs and thus the discretization errors.

It is interesting to ask
what happens to the relationship between methods (i) and (ii) 
for negative $\mu_0$.
In this case, it follows from eq.~(58) of Ref.~\cite{ShWu2} with $\omega=-\pi/2$
that the figure is symmetric under reflection in the $m_0$ axis. In fact, this
follows from the invariance of the underlying lattice theory under parity
 transformation
combined with a change of sign of $\mu_0$. Thus the line determined by
the condition (\ref{eq:PCACcond}) appears to have a cusp at the $\mu_0=0$ axis. 
How this extends into the Aoki regime is one of the questions
addressed below in sec.~\ref{sec:aoki}. 

Knowing the combination of LECs appearing in $\delta\omega$ allows one to
predict other quantities using \tmchpt. In particular, 
one of the parity violating axial form factors of
the pion [the $p_1$ term in eq.~(91) of Ref.~\cite{ShWu2} 
at maximal twist and for $q^2\to0$] depends only on $\delta_W$.
Thus, as long as one knows $Z_P/Z_S$, one can predict
this form factor up to NNLO corrections.

\bigskip
I now turn to my second observation about the relationship between methods
for determining maximal twist.
A third definition of the twist angle is given in Ref.~\cite{ShWu2},
and denoted $\omega_P$. It is based on requiring that 
the physical scalar density does not create the
neutral pion from the vacuum. Method (iii) for determining maximal
twist is defined by setting $\omega_P=\pi/2$, and is equivalent
to requiring that $\langle 0|P^3|\pi^0\rangle=0$, where $P^3$ is
the neutral pseudoscalar density in the twisted basis.
In Ref.~\cite{ShWu2} it was argued that this method would be difficult 
to implement in practice, 
since the required matrix elements involve disconnected 
quark contractions.
My observation here (which 
follows from eq.~(98) of Ref.~\cite{ShWu2}) is
that at maximum twist [defined using any of methods (i-iii)] 
\begin{equation}
\omega_P = \omega_0 + O(a^2) = \pi/2 + O(a^2) \qquad \textrm{(maximal twist).}
\label{eq:omegaPvs0}
\end{equation}
In other words, up to NNLO corrections, methods (ii) and (iii) are
the same. Thus it turns out not to be difficult to implement
method (iii) in the GSM regime.
The generalization of this result to the Aoki regime
is discussed below.

\section{\label{sec:fits} Comparing \tmchpt\ expressions to simulations}

In this short section I point out that existing simulation data
exhibit qualitative features predicted by \tmchpt,
and that by doing more detailed fits one could over-determine
the LECs associated with discretization errors.
I illustrate these points by showing plots of physical quantities
for a choice of LECs roughly consistent with the
numerical results of Ref.~\cite{Farchioni04B}.
For most of this section I work in
the GSM regime, although the plots extend into the
Aoki regime, and use results from the following section.
I assume that an optimal critical mass $m'_c$ has
been determined, i.e. that one knows the critical mass with
errors of $O(a^3)$. I then consider the functional dependence
of physical quantities on the untwisted quark mass $m'$
at fixed values of the twisted quark mass $\mu$.

As explained in Ref.~\cite{FR1}
(and extended into the GSM regime in Refs.~\cite{AB04,ShWu2,FMPR})
 $O(a)$ errors in physical
quantities can be canceled by ``mass averaging'', i.e.
by suitably averaging the results at $(m',\mu)$ and $(-m',-\mu)$
(which corresponds to averaging $\pm m_q$ 
in the physical basis used in Ref.~\cite{FR1}).
Here the term ``suitable'' indicates that, for quantities
with negative $R_5$ parity (defined in Ref.~\cite{FR1}), 
the term at $(-m',-\mu)$ must be
included with a negative sign~\cite{FR1}. Physical masses
and matrix elements, however, have positive $R_5$ parity.
One can change the average
to one between $(m',\mu)$ and $(-m',+\mu)$ by doing, for example,
a parity transformation on the second term. This transformation,
which I refer to as ``twisted parity'', 
is equivalent to $\omega\to-\omega$ in the physical basis. 
As discussed in Ref.~\cite{FR1}, it
does not effect physical quantities, but can lead to a sign change
in certain correlation functions. 
The net result is that for physical quantities one cancels $O(a)$
errors by averaging over $(m',\mu)$ and $(-m',\mu)$ with relative
positive signs, while for
other quantities one must determine the relative sign of the two
terms on a case-by-case basis. Two examples used below are
the PCAC mass and the expectation value $\langle P^3\rangle$.
The former has odd $R_5$ parity, but even twisted parity (at $\vec p=0$),
and so requires a relative minus sign for the two contributions.
The latter has odd $R_5$ and twisted parities, and so the average
has relative positive signs.

The point I want to make here is that the contributions  that are
canceled by the mass average described above are also of interest,
for they provide a measure of discretization errors, and allow
tests of \tmchpt\ at NLO.
These non-continuum parts can be picked out using
\begin{equation}
AS(Q) = \frac{Q(m',\mu)- (-)^{p}Q(-m',\mu)}
             {Q(m',\mu)+ (-)^{p}Q(-m',\mu)}
\,,
\end{equation}
where $Q$ is the quantity of interest,
and $(-)^p$ is the product of its $R_5$ and twisted parities.
The name refers to ``antisymmetric'', since for most quantities
(all those considered here except $m_{\rm PCAC}$)
$AS$ is the antisymmetric part.
Using the results of Ref.~\cite{ShWu2}, I find
\begin{eqnarray}
AS(m_{\pi^\pm}^2) &=& 2\ AS(\langle 0 | P^\mp|\pi^\pm\rangle)
= AS(\langle \pi| S^0 | \pi \rangle) 
= (m'/m_q) (2 \delta_W-\delta_{\widetilde W}) + O(a^2)
\,,\nonumber\\
AS(f_A) &=& (m'/m_q)\, \delta_{\widetilde W}/2 + O(a^2)
\,,\nonumber\\
AS(m_{\rm PCAC}) &=& (m_q/m')\, \delta_W + O(a^2)
\,,\label{eq:AS}\\
AS(\langle P_3 \rangle) &=& O(a^2)
\,,\nonumber\\
AS(m_{\pi^0}^2-m_{\pi^\pm}^2) &=& O(a)
\,.\nonumber
\end{eqnarray}
Here $f_A$ is the usual pion decay constant, and $m_{\rm PCAC}$
is defined in the following section.
I stress that these results hold in the GSM regime, 
but not in the Aoki regime.
One sees that \tmchpt\ relates the ``antisymmetries'' in the various
quantities. In particular, if $\delta_W$ has been determined
using the angle $\delta\omega$ as described in the previous section,
then only one additional parameter ($\delta_{\widetilde W}$)
is needed to describe all seven quantities in eq.~(\ref{eq:AS}).
This assumes that the ratio $Z_P/Z_S$ is known so that bare mass ratios
can be converted into those of physical masses.

One peculiar feature of the results in (\ref{eq:AS}) is the appearance
of $m_q/m'=1/\cos\omega_0$ in the result for $m_{\rm PCAC}$.
By contrast, all the other results that are non-vanishing at NLO are
proportional to $\cos\omega_0$, and thus vanish when $m'=0$.
The divergence at $m'=0$ in $AS(m_{\rm PCAC})$ results from the fact that the
average is not taken with respect to the position at which $m_{\rm PCAC}$
vanishes (which, as seen in the previous section, occurs along a line
with $\omega_0=\pi/2-\delta\omega$ and not along the line $\omega_0=\pi/2$).
In any case, the result becomes invalid when $\cos\omega_0$ becomes of
$O(a)$, i.e. when one enters the Aoki regime, so that one does not 
in fact reach the divergence.

Figure~5 of Ref.~\cite{Farchioni04B} shows plots of $m_{\pi^\pm}^2$ and
$m_{\rm PCAC}$ as a function of $m_0$ at $\mu_0=0$ and $0.01$. These results
are from dynamical simulations with a lattice spacing of
$a^{-1}\approx 1\;$GeV. In physical units (and assuming that $Z_S\approx Z_P\approx 1$)
the twisted quark mass is thus $0$ and $10\;$MeV, while $m'$ lies in the range
$ -100 \;{\rm MeV} \ltapprox m' \ltapprox 50\;$MeV. Since $a\LQ^2\approx 90\;$MeV
and $a^2\LQ^3\approx 30\;$MeV, the results lie in both Aoki and GSM regimes.
Two features of these results are relevant here. First, there is a clear
lack of symmetry in the results for $m_{\pi^\pm}^2$,
and of antisymmetry in $m_{\rm PCAC}$,
i.e. the quantities $AS(m_{\pi^\pm}^2)$ and $AS(m_{\rm PCAC})$ are 
non-vanishing. Thus it would be interesting to do a more detailed fit
to determine them. Second, since a significant fraction of the data points lie
in the Aoki region, as shown by the evidence for a first order phase transition,
a comprehensive fit requires a functional form that is valid in both GSM and Aoki
regimes. The expressions in eq.~(\ref{eq:AS}) do not suffice,
for they are valid only in the GSM regime.

The following section is devoted to providing functional forms valid in
both regimes. Since the ``antisymmetry'' arises at NLO in the GSM regime,
a consistent form must be valid also at NLO in the Aoki regime.
In fact, once one works at NLO, it may be that the best approach is
to simply fit to the full formulae for both positive and negative quark masses
rather than use the ``antisymmetries'' directly.

I close this section by illustrating the forms that result from the
joint NLO analysis in GSM and Aoki regimes. I use the following parameters,
which are roughly chosen to match the behavior seen in
Ref.~\cite{Farchioni04B}. The LECs describing discretization
errors are set to $\delta_W=\delta_{\tilde W}=-0.3$,
$2 w' = (0.25\; {\rm GeV})^2$, and $W_{3,3}=0$ ($W_{3,3}$ being
the additional non-trivial LEC needed at NLO in the Aoki region,
as discussed in the following section). In addition I set the
continuum LECs (the $L_i$ of Gasser and Leutwyler) to zero and drop chiral
logarithms. Although both these contributions are of NLO in the GSM region, 
and thus should be included for consistency, they
do not contribute to the ``antisymmetries''. Furthermore, they are of NNLO
in the Aoki regime.
Note that if I define scales by $\delta_W=-a\Lambda_W$, 
$\delta_{\widetilde W}=-a\Lambda_{\widetilde W}$,
and $2 w' = a^2 \Lambda_{w'}^4$, and use $a^{-1}=1\;$GeV,
the parameters I use
correspond to $\Lambda_W=\Lambda_{\widetilde W}=300\;$MeV,
and $\Lambda_{w'}=500\;$MeV, which are reasonable values for the
scales of discretization errors. The choice $W_{3,3}=0$ is made for simplicity.
Taking $w'>0$ implies that one is in the first-order scenario
of Ref.~\cite{ShSi}, consistent with the numerical results.
The quantity $\sqrt{2 w'}=250\;$MeV is then approximately the
minimum pion mass. See Fig.~\ref{fig:phases:b} for a sketch of the
phase diagram.

\begin{figure}
\centering
\subfigure[]{
\label{fig:mpi:a}
\scalebox{1}[1]{\includegraphics[width=3in]{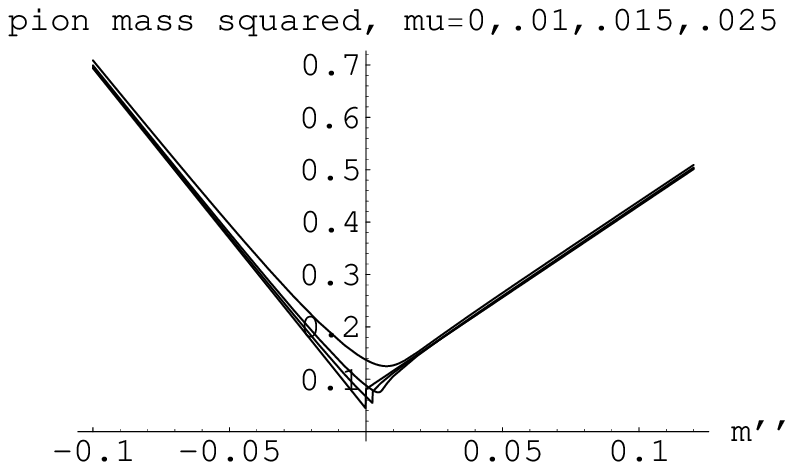}}}
\hspace{0.5in}
\subfigure[]{
\label{fig:mpi:b}
\scalebox{1}[1]{\includegraphics[width=3in]{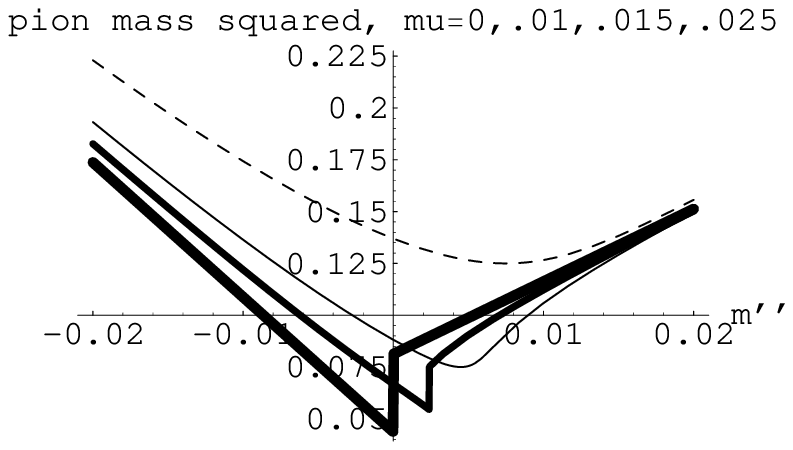}}}
\caption{\label{fig:mpi}
$m_{\pi^\pm}^2$ as a function of $m''$ for fixed $\mu=0,0.01,0.015,0.025$,
for the parameter set described in the text. 
Here $m''$ is the NLO quark mass in the Aoki regime,
defined in eq.~(\protect\ref{eq:mprprdef}). For the GSM
regime [i.e. most of fig.~(a)] $m''$ is equivalent to $m'$
at NLO accuracy.
All quantities are in units of appropriate powers of GeV. 
The curves in (a) can be identified from the enlargement (b):
the lines decrease in thickness 
as $\mu$ increases, with the $\mu=0.025$ curve dashed.
}
\end{figure}

\begin{figure}
\centering
\subfigure[]{
\label{fig:PCAC:a}
\scalebox{1}[1]{\includegraphics[width=3in]{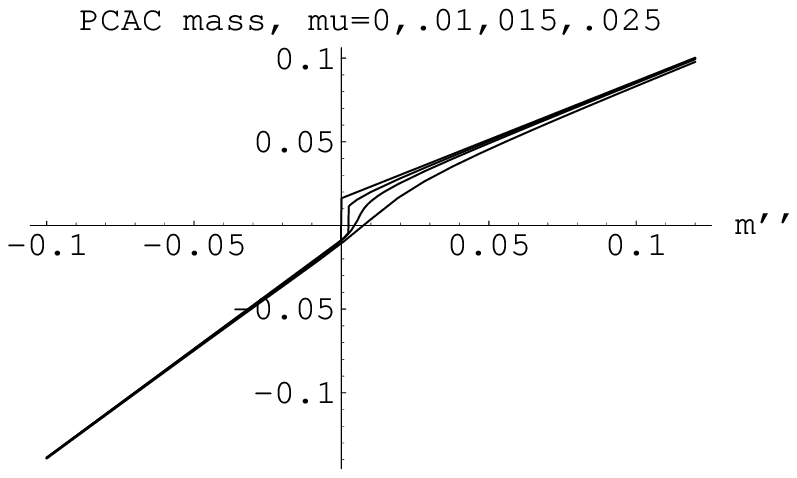}}}
\hspace{0.5in}
\subfigure[]{
\label{fig:PCAC:b}
\scalebox{1}[1]{\includegraphics[width=3in]{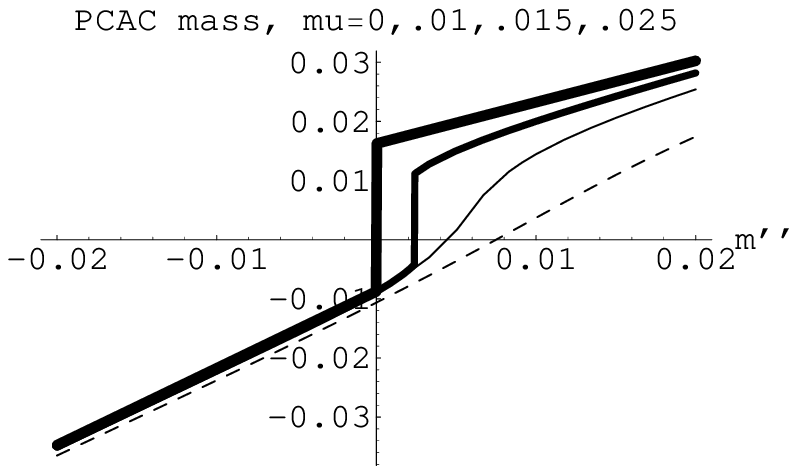}}}
\caption{\label{fig:PCAC}
$m_{\rm PCAC}$ as a function of $m''$, with parameters and legend 
as in Fig.~\protect\ref{fig:mpi}.
}
\end{figure}

Figure~\ref{fig:mpi} shows the forms of $m_{\pi^\pm}^2$ for
$\mu$ fixed at $0$, $10$, $15$ and $25\;$MeV, and Fig.~\ref{fig:PCAC}
shows the corresponding plots for $m_{\rm PCAC}$. In both figures
the left-hand plot shows the results primarily in the GSM regime,
while the right-hand plot zooms in on the Aoki region. 
Within the GSM regime, the mass $m''$ (defined in the
following section) is equivalent, at NLO, to the
mass $m'$ used in this section. The prediction of
eq.~(\ref{eq:AS}) that the ``antisymmetries'' should be independent
of $\mu$ holds to good approximation for $|m''|\gtapprox 50\;$MeV. 
The values of $AS$ for both quantities asymptote at large $|m''|$ to $-0.3$, 
roughly the values one obtains from the data of Ref.~\cite{Farchioni04B}.

I discuss the features in the Aoki regime in the next section.
I only note here that the first order
phase transition can be seen in the jumps for $\mu=0$ and $10\;$MeV.
The end-point of the transition is for $\mu=w'/B_0=12.5\;$MeV
(as discussed in the following section) so that the lines with
$\mu=15$ and $25\;$MeV pass above the transition. 

\section{\label{sec:aoki} The Aoki regime at NLO}

In this section I extend the \tmchpt\ analysis in the Aoki region
to NLO. The motivation for making this
extension is twofold. First, it allows one to see what happens in
detail to the phase transition line, and to the lines of
maximal twist, as one enters the Aoki region. 
In particular, it
clarifies the size and parametric form of the errors that
are made in different schemes for determining maximal twist.
Second, it provides NLO results that can be used consistently
in the Aoki and GSM regimes.

\begin{figure}
\centering
\subfigure[Aoki-phase]{
\label{fig:phases:a}
\scalebox{1}[1]{\includegraphics[width=3in]{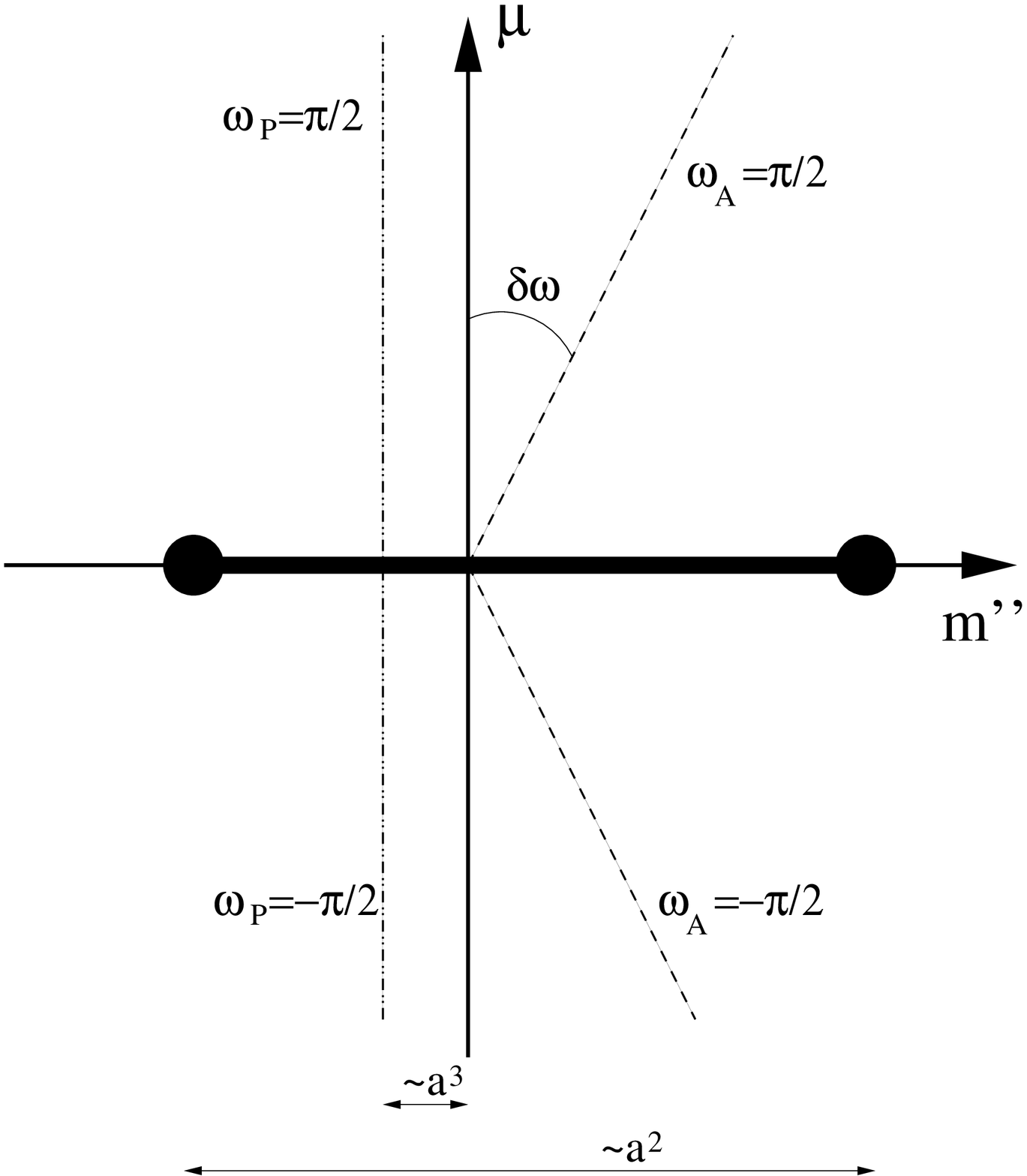}}}
\hspace{0.5in}
\subfigure[First-order]{
\label{fig:phases:b}
\scalebox{1}[1]{\includegraphics[width=3in]{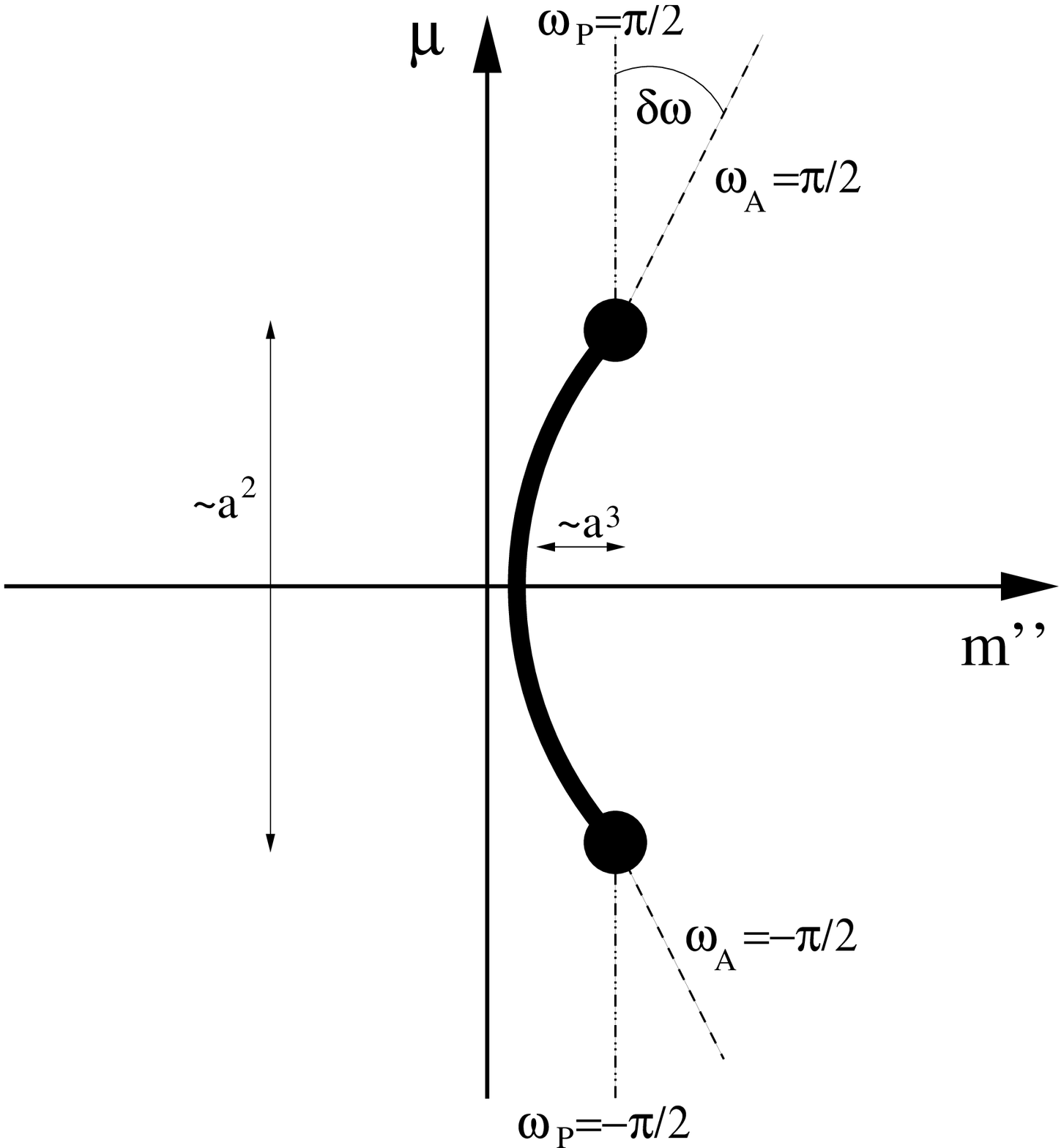}}}
\caption{\label{fig:phases}
Illustration of phase structure and lines of maximal twist
at NLO in the Aoki regime for both Aoki-phase and first-order scenarios.
The solid lines indicates first-order transition lines, 
with the solid circles being second-order end-points.
Both masses $m''$ and $\mu$ range up to values of $O(a^2)$ in this figure,
with other features (the asymmetry in the phase line in the Aoki-phase,
the offset of the $\omega_P=\pm\pi/2$ lines, and the horizontal extent
of the phase-line in the first-order scenario) being of $O(a^3)$.
See the text for more details. The angle $\delta\omega$ is the
same as that in Fig.~\protect\ref{fig:maxtwist}.}
\end{figure}

The main results of this section are that a NLO analysis in the Aoki regime
requires only one non-trivial additional LEC, and that this leads to
changes in the phase diagram, as summarized in Fig.~\ref{fig:phases}.
These changes are small, proportional to $a^3$ as compared to the
size of the phase boundary ($\sim a^2$), but lead to new effects which might
be measurable, e.g. curvature in the phase boundary, and a discontinuity
in the pion mass across the boundary. I also 
determine the lines of maximal twist, and results
for a number of physical pionic quantities, both at NLO.
These are of theoretical
interest, and may be needed to describe the results of numerical simulations.
Finally, I describe what happens to the phase diagram if the
gauge action is tuned so as to reduce the size of the LEC which
determines the phase structure at LO from $\sim a^2$ to $\sim a^3$.

\bigskip

In the Aoki regime, LO terms are of size $m_q\sim a^2$ (terms linear in
$a$ having been absorbed into the quark mass), 
and NLO contributions are suppressed by a power of $a$.
Thus NLO terms are of size $m_q a$, $p^2 a$ and
$a^3$. The first two are also of NLO in the GSM regime,
and thus are already included in the NLO chiral Lagrangian of
Ref.~\cite{ShWu2}. The $a^3$ terms are new. Using the
properties of $SU(2)$, I find that there are two independent 
terms:
\begin{equation}
{\cal L}_\chi^{(a^3)} 
=
- \frac{W_{3,1}}{f^2}
\left[\Tr(\hat A^\dagger \hat A)\Tr(\hat A^\dagger \Sigma) + h.c. \right]
- \frac{W_{3,3}}{f^2} 
\left[\Tr(\hat A^\dagger \Sigma)^3 + h.c.\right]
\,,
\end{equation}
where $h.c.$ indicates hermitian conjugate,
and $\hat A$ is the spurion associated with discretization errors,
which is set to $\hat a$ times the identity matrix.
The new LECs, $W_{3,1}$ and $W_{3,3}$, are dimensionless.

Recalling the form of the leading order mass term
\begin{equation}
-\frac{f^2}{4}\Tr(\chi'^\dagger \Sigma)+ h.c.
\,,
\end{equation}
where $\chi'=\hat{m}'+ i\tau_3 \hat\mu$ is the mass spurion,
we see that the $W_{3,1}$ term can be absorbed
by a redefinition of $\chi'$ and $\hat{m'}$:
\begin{equation}
\chi' \longrightarrow \chi'' = \chi' + \frac{8 \hat{a}^3 W_{3,1}}{f^4}
\equiv \hat{m}'' + i \tau_3 \hat\mu
\,.
\label{eq:mprprdef}
\end{equation}
This is equivalent to an $O(a^3)$ shift in $m_c$. Since one does not
know $m_c$ {\em a priori}, this shift does not increase our ignorance.
All the formulae quoted above, and in Ref.~\cite{ShWu2},
hold to their stated accuracy with $\chi'$ replaced by $\chi''$ 
(and thus $M'$ replaced with $M''= \sqrt{{\hat{m}''}{}^2+\hat\mu^2}$
and using $\tan\omega_0=\mu/m''$).

Thus there is only one non-trivial new term to include in the analysis.
Before describing how this changes the LO results, I want to
comment on redundancy in the LECs. 
In Ref.~\cite{ShWu2}, it was shown that the LEC $W_{10}$ could be
set to zero by a change of variables, as long as other LECs were
adjusted accordingly. This implied that, if one kept $W_{10}$
in the analysis, the results would depend only on certain linear
combinations of LECs. Extending this analysis to the $O(a^3)$ terms,
I find that the contributions of the 
two new constants to physical quantities
must appear in
combinations $2 W_{3,3}+W_{3,1}$ and $W_{3,3}- 32 W' W$.
I have used this to check the results given below, but,
to simplify expressions, I have set $W_{10}$ to zero having
made these checks.

Now I turn to the results.  My aim is to map out the phase structure 
at fixed $a$ as a function of $m''$ and $\mu$,
and to determine NLO expressions for physical quantities throughout
this mass plane. Note that, although $m''$ is not known
{\em a priori}, 
one can use the results given here to determine $m''$
{\em a posteriori}. 
Where appropriate I will include contributions which,
while of NNLO in the Aoki regime, are of NLO in the GSM regime, 
and thus are needed for a complete, continuous description within
the mass plane. 

I begin with the determination of the expectation
value, $\Sigma_m$, of the pion field $\Sigma\in SU(2)$.
This is the value which minimizes the classical potential,\footnote{%
Loop effects will be shown below to be of NNLO.}
which is here
\begin{equation}
\frac{V_\chi}{f^2} = -(c_m \hat{m}'' + s_m \hat\mu)(1 + c_m \delta_W)
- c_m^2 w' - c_m^3 w_3 + O(a^4)
\,,
\end{equation}
where I use 
\begin{equation}
c_m = \Tr(\Sigma_m)/2\,,\ \ \textrm{and}\ \ 
s_m = -i\Tr(\Sigma\tau_3)/2\,,
\end{equation}
and define the useful quantity
\begin{equation}
%\delta_W = \frac{16 \hat{a} W}{f^2}\,,\ \
%w' = \frac{16 \hat{a}^2 W'}{f^2}\,,\ \
w_3 = \frac{16 \hat{a}^3 W_{3,3}}{f^4}\,.
\end{equation}
The range of the variables is $-1 \le c_m, s_m \le 1$,
with the constraint $-\sqrt{1-c_m^2}\le s_m \le \sqrt{1-c_m^2}$. 
For $\hat\mu\ne 0$, the linear dependence of $V_\chi$ on $s_m$ implies
that $s_m$ is pushed to one or other end of its allowed range, so that
$s_m^2+c_m^2=1$. This implies that $\Sigma_m$ lies 
in the $\tau_3$ direction.
For $\hat\mu=0$
(the Wilson axis), one can choose the expectation value to
lie in the $\tau_3$ direction. Thus from now on I set
$\Sigma_m = \exp(i\omega_m \tau_3)$, so that
$s_m=\sin\omega_m$ and $c_m=\cos\omega_m$.

In the Aoki region, the LO potential, obtained by setting
$\delta_W=w_3=0$, has been analyzed in Refs.~\cite{Mun04,Scor04,ShWu1}.
There are two scenarios depending on the sign of $w'$, as
shown (including NLO effects) in Fig.~\ref{fig:phases}. 
For $w'<0$, a transition line lies along
the Wilson axis, within which parity and flavor are spontaneously
broken (the Aoki phase~\cite{Aoki}), with endpoints at 
$\hat{m}''=\mp 2 w'$. This I call the ``Aoki-phase scenario''.
For $w'>0$,
there is a first order transition line along the $\mu$ axis with
second-order endpoints at $\hat\mu=\pm 2 w'$. This I call the
``first-order scenario''. 

The extension of this analysis to NLO is straightforward in principle.
The extra terms are, by assumption, small, and do not change the
qualitative features of the two scenarios. They do, however, lead to
small changes in the positions of the transition 
lines and the predictions for other quantities.

I consider first the Aoki-phase scenario ($w'<0$). 
Along the Wilson axis, the $-c_m \hat{m}''$ term in the potential
ensures that $c_m$ is forced to one or other end of its
allowed range for large enough $|\hat{m}''|$: $c_m = {\rm sgn}(\hat{m}'')$.
The phase transition occurs when the other terms cancel this force,
i.e. when  $d V_\chi/d c_m (c_m=\pm 1) = 0$. This occurs at
\begin{equation}
\hat{m}'' = \mp 2 w' - 3 w_3 + 4 w' \delta_W + O(a^4) 
\qquad \textrm{(Aoki-phase endpoints)}\,.
\label{eq:Aokiphaseendpoints}
\end{equation}
The first term on the r.h.s. is the LO result, which is of $O(a^2)$.
The next two terms are the $O(a^3)$ corrections, which lead 
 to a shift in the phase transition line,
but do not change its length. Of course, to measure this shift
requires knowing where the origin is, and I return to this point
below. Within the Aoki phase, 
the identity component $c_m$ varies continuously from $+1$ to $-1$ 
(in a way which can be determined, if needed, by minimizing $V_\chi$).
I only note here that the expectation
value points purely in $\tau_3$ direction ($c_m=0$) when
$\hat{m}''=0$, which is {\em not} the mid-point of
the transition line (unlike at LO).
As at LO, the charged pions are massless in the Aoki phase,
with the neutral pion also massless at the end points.

The changes at NLO are more significant in the first-order scenario.
There is no symmetry which requires the transition to lie parallel to
the $\mu$ axis once NLO terms are included.
To begin I consider the position of the transition line on the Wilson
axis. The transition occurs when the two minima at $c_m=\pm 1$
have equal energy. Because the $w_3$ term in $V_\chi$ is an odd
function of $c_m$,
this occurs not at $\hat{m}''=0$ (the LO result)
but at $\hat{m}''= - w_3$. For $\hat{m}''$ greater than (less than)
this value, $c_m=+1$ ($-1$).

This analysis is easily generalized off the Wilson axis.
At LO, the first-order
transition is present at $\hat{m}''=0$ if $|\hat\mu|< 2 w'$.
The LO potential has equal minima at
$c_m=\pm c_m^0 = \pm \sqrt{1-(s_m^0)^2}$,
both having $s_m=s_m^0=\hat\mu/(2w')$. The two minima coalesce
into second-order endpoints at $\hat\mu=\pm2w'$.
At NLO, the terms odd in $c_m$ give opposite contributions
at the two minima, and one must adjust
$\hat{m}''$ to make their energies equal.
To the order I work, $\hat{m}''$ can be determined
by enforcing $V_\chi(c_m=c_m^0) = V_\chi(c_m = -c_m^0)$.
From this it follows that the transition is shifted to
\begin{equation}
\hat{m}'' = - w_3 +
 (w_3- 2 w' \delta_W) \left(\frac{\hat\mu}{2 w'}\right)^2 + O(a^5)
\qquad
(\textrm{1st order line:}\ |\hat\mu|\le 2w')
\,.
\label{eq:firstorderline}
\end{equation}
The transition line is thus a quadratic function of $\mu$,
which maintains the symmetry under $\mu\to-\mu$ while being smooth
at $\mu=0$. The value of $\hat\mu$ at the end-points
is unchanged from LO, aside from NNLO corrections of $O(a^4)$
discussed below. For a given $\hat\mu$, however, the error in
$\hat{m}''$ turns out to be of $O(a^5)$,
rather than the naively expected $O(a^4)$.
Finally I note that the NLO terms shift the angle $\omega_m$ at the
transition line from the LO value by
\begin{equation}
\delta\omega_m = \omega_m-\omega_m^0=(\delta_W - w_3/w') (\hat\mu/[2 w'])+O(a^2)
\,.
\label{eq:deltaomegam}
\end{equation}
This $O(a)$ shift is in the {\em same direction} on both
sides of the transition, so the resulting values of 
$c_m$ on the two sides differ
in magnitude: $c_m=\pm c_m^0 - (\delta\omega_m )\,s_m^0$.
This leads to the discontinuity in $m_\pi^2$ at the transition,
as shown in Fig.~\ref{fig:mpi:b} and discussed below.

To understand the size of the uncertainty in the position
of the end-points it is useful to display the form of the potential
there. The considerations are identical
for both end-points, and I choose to consider that
at positive $\mu$. Setting $\hat\mu=2w'$ and
$\hat{m}''=-2 w' \delta_W$ [from eq.~(\ref{eq:firstorderline})], one has
\begin{equation}
\frac{V_\chi}{f^2} = \frac{w' c_m^4}{4}
 + (w'\delta_W - w_3) c_m^3 + 2 w' \delta_W^2 c_m^2
\qquad
\label{eq:Vatendpoint}
\textrm{(1st order endpoint).}
\end{equation}
I have dropped terms of higher order than quartic in $c_m$, as they would
be subleading in the following discussion.
At LO, only the first term on the r.h.s. contributes.
It has a minimum at $c_m=0$, and
 corresponds to the end-point since the curvature vanishes.
At NLO, the value of $\hat{m}''$ has been chosen by the criterion
given above so that $c_m=0$
remains a stationary point of the potential. 
$c_m=0$ does not, however, correspond
to an end-point, for there is a non-vanishing quadratic term at $c_m=0$
(with coefficient $\propto a^4$). Furthermore, for some values of the LECs, 
the cubic term leads to a second stationary point with $c_m\sim a$.
To move to the actual end-point, one must shift $\hat\mu$ and, in general,
$\hat{m}''$. For example,
the quadratic term can be canceled by a shift of size $\delta\hat\mu \sim a^4$,
because $\delta V_\chi \propto \delta\hat\mu (1-c_m^2/2 + \dots)$.
It can be shown that a similarly sized shift in $\hat\mu$,
as well as a shift $\delta\hat{m}''\sim a^5$,
will also remove the second minimum, if present.

A NLO calculation does not, however, control the size of these shifts.
This is because there are neglected higher order contributions to the potential which 
are of the same size as those in eq.~(\ref{eq:Vatendpoint}) when $c_m\sim a$.
For example, there will be NNLO and higher order 
contributions to the classical potential of the form 
$\delta V_\chi \sim a^4 c_m^2 + a^5 c_m$, depending on
additional unknown LECs. These, like the terms in
(\ref{eq:Vatendpoint}), are of $O(a^6)$ when $c_m\sim O(a)$,
and thus compete with NLO terms. In particular, the
term linear in $c_m$ leads to an unknown shift $\delta\hat{m}''\sim a^5$,
and the quadratic term to a shift $\delta\hat\mu\sim a^4$.

In addition, loop contributions to the effective potential 
enter at the same order. Flavor-conserving four-pion vertices of LO
(and thus proportional to $p^2$ or $m_q$) give rise to quadratic terms
in the potential of the form $\delta V_\chi \sim c_m^2 m_\pi^4 \ln(m_\pi)
\sim a^4 c_m^2$. The latter equality follows because,
for charged pions, $m_\pi^2 \sim a^2$
at the end-points.
There are also flavor-violating four-pion vertices
of the form $a^2 s_m^2 W' \vec \pi\cdot \vec \pi \pi_3^2$ (see Ref.~\cite{ShWu2}),
and these also lead to 
$\delta V_\chi\sim c_m^2 a^2 m_\pi^2 \ln m_\pi\sim a^4 c_m^2$.
Finally, there are tadpole diagrams involving three-pion vertices. 
These vertices have been given in Ref.~\cite{ShWu2} in the GSM regime. 
In the Aoki regime they become
\begin{equation}
f {\cal L}_\chi^{3\pi} = 
- \delta_{\tilde W}\pi_3 \partial_\mu \vec \pi\cdot\partial_\mu \vec\pi
+ \frac{s_m \hat{m}'' - c_m \hat\mu}{2} \pi_3 \vec \pi \cdot \vec \pi
\,.
\end{equation}
Noting
that the coefficient of the second term is of $O(a^3)$ at the end-points, 
I find that the contributions to the potential are of the form
\begin{equation}
\delta V_\chi^{\rm tad}\sim c_m (a m_\pi^4\ln m_\pi
+ a^3 m_\pi^2\ln m_\pi) \sim a^5 c_m
\,,
\end{equation}
and thus of the same size as the higher order terms in the classical
potential for $c_m\sim a$.
Clearly extending the calculation beyond NLO will be challenging.\footnote{%
The breakdown of power-counting near the end-points has been noted
in the Aoki-phase scenario in Ref.~\cite{Aoki03}.
}

A corollary of the previous discussion is that, close to the end point,
the NLO calculation does not control the shift $\delta\omega_m$, since
higher order terms contribute with the same size as LO and NLO terms.
For the result (\ref{eq:deltaomegam}) to be valid,
the distances from an end-point must
satisfy $\delta\hat\mu \gg a^4$ or $\delta\hat{m}'' \gg a^5$.
Furthermore, as one approaches the end-points the rate of convergence
of the chiral expansion slows.
For example, if one moves a distance 
$\delta\hat{m}''\sim a^3$ from the end-point, then one can show that
$c_m \sim a^{1/3}$. It follows that each higher order is suppressed by
a relative factor of $\sim a^{2/3}$, rather than $\sim a$.

The results for the phase structure are summarized
in Fig.~\ref{fig:phases}. The circles around the end-points
emphasize that the chiral expansion breaks down in their vicinity,
as just discussed.
Note that, since the origin of $\hat{m}''$ is
not known, in the Aoki-phase scenario one can, at this stage, only
determine $w'$ from the length of the phase region, while in the
first-order scenario one can determine both $w'$ (from the length in
the $\mu$ direction), and $w_3-2 w'\delta_W$ (from the extent
of the phase line in the $\hat{m}''$ direction).

\bigskip
I now turn to results for other physical quantities. To do so,
one needs to know the value of $\omega_m$ for a general position
in the mass plane. This is determined by solving
\begin{equation}
0 = \frac{1}{f^2}\frac{d V_\chi}{d\omega_m} =
s_m \hat{m}'' - c_m \hat\mu + 2 c_m s_m w'
+(2 c_m s_m \hat{m}'' + [s_m^2-c_m^2] \hat\mu)\delta_W
+ 3 c_m^2 s_m w_3 + O(a^4)
\,,
\label{eq:minimize}
\end{equation}
and choosing the solution with minimal energy.
In practice, for general values of the parameters,
and away from the phase boundaries,
one can solve the LO equation (the first three terms on
the r.h.s.) and use this solution as the starting estimate
in a numerical solution to the full NLO equation.

I first consider the lines along which $\omega_A=\pm \pi/2$,
i.e. the lines of maximal twist as defined by method (i) of 
sec.~\ref{sec:maxtwist}. The advantage of choosing variables
such that $W_{10}=0$
is that $\omega_A=\omega_m+ O(a^2)$ \cite{ShWu2}.
Thus the task is to solve for $\omega_m = \pm \pi/2$, so that $c_m=0$
and $s_m=\pm 1$. For these values,
eq.~(\ref{eq:minimize}) simplifies considerably, leading to
\begin{equation}
\hat{m}'' = \left[\mp \delta_W + O(a^2) \right]\hat\mu
\qquad
(\omega_A=\pm\pi/2).
\label{eq:omegaAmax}
\end{equation}
In the Aoki-phase scenario, this result holds for all $\hat\mu$, 
while in the first-order scenario it holds for $|\hat\mu|$ down to
a minimum value of $2 w'$, at which point the $\omega_A=\pm \pi/2$ lines
run into the second-order end-points.\footnote{%
There is an apparent inconsistency here: the end-points both have
$\omega_m=\omega_A=\pm\pi/2$ and, according to eq.~(\ref{eq:deltaomegam}),
$\omega_m=\pi/2 + \delta\omega_m$. This is resolved by the fact
that the $O(a)$ corrections to $\omega_m$ are not controlled in
the present calculation near the end-points.}
For smaller $\hat\mu$ one can show
that the solution (\ref{eq:omegaAmax}) does not have minimum energy.
The situation is shown in Fig.~\ref{fig:phases}.

The result eq.~(\ref{eq:omegaAmax}) answers the question
(posed in sec.~\ref{sec:maxtwist}) of what happens to the lines of
maximal twist when one enters the Aoki region. The answer is that they
remain straight [the angle given in eq.~(\ref{eq:deltaomega})
agrees with that in eq.~(\ref{eq:omegaAmax}) to the stated accuracies],
and extrapolate to the origin in the mass plane
as long as one uses the variable $m''$ rather than $m'$.
In other words, the $O(a^3)$ uncertainty in $m'$ in the GSM regime
(discussed in sec.~\ref{sec:maxtwist}) is resolved by
extending the calculation into the Aoki regime.

Note that in both phase-transition scenarios one 
can determine $\delta_W$, $w'$ and $w_3$ from knowledge of the
position of the phase transition lines and of the line on
which $\omega_A=\pi/2$. In both cases, $\delta_W$ comes from the slope
(assuming $Z_S/Z_P$ is known), and $w'$ from the extent of the phase
transition line. In the Aoki-phase scenario, the phase line is asymmetric
with respect to the origin, and from this one can determine $w_3$
using eq.~(\ref{eq:Aokiphaseendpoints}).
In the first-order scenario, one can use the curvature of the phase boundary
or, equivalently, its intercept with the Wilson axis, to determine $w_3$,
using eq.~(\ref{eq:firstorderline}).

I next determine the lines along which $\omega_P=\pm\pi/2$.
It is shown in Ref.~\cite{ShWu2} that $\omega_P = \omega_m + s_m\delta_W + O(a^2)$
in both GSM and Aoki regimes. Thus one needs to solve eq.~(\ref{eq:minimize})
for the lines along which $\omega_m = \pm \pi/2 \mp \delta_W$.
At the accuracy I work this means that $s_m= \pm 1$ and $c_m=\delta_W$.
The result is very simple:
\begin{equation}
\hat{m}'' = - 2 w' \delta_W + O(a^4)
\qquad
(\omega_P=\pm\pi/2).
\label{eq:omegaPmax}
\end{equation}
Recall that in the GSM regime, the $\omega_P=\pm\pi/2$ line lies on the
$\mu$ axis for any suitably accurate definition of the critical mass.
In particular, one could extrapolate the $\omega_A=\pm \pi/2$ lines to
$\mu=0$ and use this to define $m_c$ [i.e. method (ii) of sec.~\ref{sec:maxtwist}].
The result (\ref{eq:omegaPmax}) shows that, 
with the greater resolution provided by working at NLO in the Aoki regime, there
is an $O(a^3)$ offset between the definitions of $m_c$ determined from $\omega_A$
and $\omega_P$. In other words, within the Aoki regime, methods (ii) and (iii)
differ at NLO.

The situation is illustrated in Fig.~\ref{fig:phases}. 
Note that, in both scenarios, the position 
of the $\omega_P=\pi/2$ line is a prediction of \tmchpt, since it depends
on the combination $\delta_W w'$ that can be determined as discussed above.
In the Aoki-phase scenario, the lines of $\omega_A=\pi/2$ and $\omega_P=\pi/2$ do not
cross for either sign of $\delta_W$. 
In the first-order scenario, the lines do meet, and they do so at the end-points of
the phase transition line. In fact, since the $\omega_m=\pi/2$ solution ends at
these points, so do the $\omega_P=\pi/2$ lines.

I have focused on the lines of maximal twist since these are of
greatest practical interest. I note, however, that one can predict the
values of $\omega_A$ and $\omega_P$ to NLO accuracy using the formulae
given above throughout the Aoki and GSM regimes. In particular, the
jump in $|c_m|$ as one crosses the first-order line 
[see eq.~(\ref{eq:deltaomegam}) and subsequent discussion]
applies also to $|c_A|$, since $\omega_A=\omega_m$. 

\bigskip
I now turn to pion properties. I find the charged pion masses
to be
\begin{equation}
m_{\pi^\pm}^2 = (c_m \hat{m}'' + s_m \hat\mu)(1 + c_m[2 \delta_W-\delta_{\widetilde W}])
+ 2 c_m^2 w'  + c_m^3 (3 w_3-2 \delta_{\widetilde W} w')
+ M''^2\ {\rm terms} + O(a^4) \,.
\label{eq:chargedpionmassAoki}
\end{equation}
The ``$M''^2$ terms'' are as in the continuum, with $2B_0 m_q$ replaced by
$M'' = \sqrt{\hat{m}''{}^2+\hat\mu^2}$. They include chiral logarithms,
and are given, for example, in Ref.~\cite{GLoneloop}. 
Within the Aoki regime they are of
$O(a^4)$ and thus of NNLO, but they are of NLO in the GSM regime and must
be included there. The new NLO terms in the result (\ref{eq:chargedpionmassAoki})
are those proportional to $a^3 c_m^3$. Given a set of LECs, one can obtain
the pion mass at the desired position in the mass plane 
by first determining $\omega_m$ using eq.~(\ref{eq:minimize})
and then substituting in eq.~(\ref{eq:chargedpionmassAoki}). 
This is how the plots in secs.~\ref{sec:fits} and \ref{sec:IR} are made.

Although the general formula is complicated, it simplifies at maximal twist.
In particular, if one works along
the $\omega_A=\omega_m=\pi/2$ line, i.e. using 
method (i), then all the NLO corrections vanish:
\begin{equation}
m_{\pi^\pm}^2 = \hat\mu + M''^2\ {\rm terms} + O(a^4) \qquad (\omega_A=\pi/2).
\label{eq:mpimaxtwist}
\end{equation}
Thus not only are $O(a\mu)$ terms absent, but also $O(a^2)$ and $O(a^3)$ terms
vanish. The result thus takes the continuum form 
(with $m_q=\mu$, and up to $O(a^4)$ corrections)
all the way until the end of the $\omega_A=\pi/2$ line (i.e. until $\mu=0$
for the Aoki-phase scenario, and until $\mu=2w'$ for the first-order scenario).
The same result holds to the stated accuracy with both methods (ii) and
(iii) applied in the Aoki regime, aside from one caveat.
On the $\omega_P=\pi/2$ line, $c_m\sim a$ and $m''\sim a^3$, so the
terms in eq.~(\ref{eq:chargedpionmassAoki}) which differ from the continuum
result remain of $O(a^4)$ in the Aoki regime. The same is true on the
$m''=0$ line in the Aoki-phase scenario. The caveat is that, on the $m''=0$
line in the first-order scenario, $\omega_m$ starts to differ by $O(1)$ from
$\pi/2$ when $\mu$ approaches the end-point. Thus one can only use method (ii)
in this case down to $\hat\mu > 2 w'$.

Along the Wilson axis, by contrast, there are corrections of $O(a\mu)$, $O(a^2)$
and $O(a^3)$. These lead to a complicated behavior, including
a discontinuity at the first-order boundary:
\begin{equation}
m_{\pi^\pm}^2 = 2 w' \pm 2 (w_3 - \delta_{\widetilde W} w') + O(a^4)
\qquad
\textrm{(first-order discontinuity,\ }\mu=0).
\end{equation}
The behavior can be seen in Fig.~\ref{fig:mpi:b}, which also
illustrates the curvature of the transition line because the boundary
is at non-zero $m''$ for $\mu\ne 0$.

One check on the result eq.~(\ref{eq:chargedpionmassAoki}) is that,
in the Aoki-phase scenario, the charged pion mass vanishes
at the end-points, and remains zero (within the $O(a^4)$ accuracy) inside
the Aoki phase.

The NLO result for the flavor-breaking splitting 
$\Delta m_\pi^2 = m_{\pi^0}^2-m_{\pi^\pm}^2$ is
\begin{equation}
\Delta m_\pi^2 = - 2 s_m^2 w' + 
2 s_m^2 c_m (w'[\delta_{\widetilde W}+ 2 \delta_W] - 3 w_3) + O(a^4)
\,.
\label{eq:masssplitting}
\end{equation}
Note that the $a^3$ term vanishes both on the Wilson axis and
for $\omega_m=\pm \pi/2$. In fact, to NLO accuracy, $\Delta m_\pi^2= -2 w'$
along both the $\omega_A=\pm\pi/2$ and $\omega_P=\pm\pi/2$ lines,
i.e. for both methods (i) and (iii), as well as for method (ii) with
the same caveat as discussed for the charged pion mass above.

The $a^3$ terms do not contribute to the pion decay constants
$f_A$ and $f_P$, nor to the scalar form factor of the pion
or the condensates. Thus the expressions for these quantities
given in Ref.~\cite{ShWu2} remain valid and I do not repeat them here.
I do stress, however, that at maximal twist, using any of the definitions,
all these quantities agree with their continuum forms up to NNLO.

The PCAC mass does, however, receive contributions from $O(a^3)$ terms.
The definition that is useful for \tmchpt\ is
\begin{equation}
m_{\rm PCAC} \equiv \frac{\langle 0 | \partial_\mu A_\mu^1 | \pi^1 \rangle}
                         {2 \langle 0 | P^1 | \pi^1 \rangle}
\,.
\end{equation}
This leads to the general expression~\cite{ShWu2}
\begin{equation}
m_{\rm PCAC} = \frac{\mu}{\tan\omega_A} = \frac{\mu}{\tan\omega_m}[1 + O(a^2)]
\,,
\end{equation}
which is useful except on the Wilson axis, 
or to the result
\begin{equation}
m_{\rm PCAC} = \frac{c_m m_{\pi^\pm}^2 f_A}{2 f_P} [1 + O(a^2)]
= \frac{c_m m_{\pi^\pm}^2}{2 B_0}
\left[1 + c_m(\delta_{\widetilde W} - \delta_W)
+ O(M'') + O(a^2) \right]\,.
\end{equation}
These formulae are used to make the plots in Fig.~\ref{fig:PCAC}.
Note that the jump in $m_{\rm PCAC}$ at the phase-boundary
is asymmetric, i.e. the minimum value is different 
on the two sides of the transition.

\bigskip
I close this section with a brief discussion of what happens if,
by adjusting the gluon and quark actions, one were able to reduce the
size of $w'$ from $O(a^2)$ to $O(a^3)$. This would be advantageous from
a practical viewpoint since the size of the region in which lattice
artifacts have an important influence on vacuum alignment 
(the Aoki regime) would be reduced.
Furthermore, since $w'$ is the only $O(a^2)$ term, discretization
errors in general would be reduced.

The discussion of phase structure given earlier was predicated on
the $O(a^3)$ terms being small perturbations to the $w'$ term. If this is
not the case, then qualitative changes are possible.
A general analysis for $w'\sim O(a^3)$ is straightforward in principle
(although one must work only at LO, since $O(a^4)$ terms
are not controlled) but I have only worked out
in detail the simplest case of $w'=0$. The vacuum is then
determined by competition between cubic and linear functions of $c_m$. 
I find only one scenario: as one
moves along the $\mu=0$ axis there is a second-order endpoint, 
followed by a region of
Aoki-phase with spontaneous flavor-parity breaking of length
$\sim a^3$, and then by a first order
transition. This latter transition presumably extends 
in the $\pm\mu$ directions for a distance of $\sim a^3$, ending at second-order
endpoints.
In other words, the usual two scenarios are merged into one having
features of both. The remnant of the presence of two scenarios
for $w'\sim O(a^2)$ is that the relative positions of the first-order transition and
Aoki-phase segment depends on the
sign of $w_3$. Presumably, as $w'$ increases in magnitude, one or other feature
of this picture will reduce and then disappear. For example, if $w'$ is positive
the Aoki-phase segment will reduce in size and then disappear,
while the  length of the first-order transition will increase.

\section{\label{sec:IR} The absence of infra-red divergences} 

In this section I comment on the work of Ref.~\cite{FMPR}.
Based on a general analysis using the Symanzik expansion~\cite{Symanzik},
Ref.~\cite{FMPR} finds apparently infrared (IR) divergent discretization
errors in physical quantities at maximal twist which are at worst
of the form $(a/m_\pi^2)^{2k}$, with $k\ge1$ an integer. 
These divergences are interpreted as indicating the onset of
large discretization errors as $m_\pi^2$ is reduced,
and the breakdown of the
Symanzik expansion when $m_\pi^2\sim m_q < a$.
This result is for a non-optimal choice of $m_c$
having an error of $O(a)$. Using an optimal choice
(such as those discussed in sec.~\ref{sec:maxtwist}) 
Ref.~\cite{FMPR} finds the divergences to be weakened, 
such that the leading IR divergent contribution
is proportional to $a^4/m_\pi^2$.
This leads to the conclusion that it is possible, with an optimal
choice of $m_c$, to work with quark masses as small as
$m_q > a^2$, although for smaller masses, i.e. in the Aoki regime,
discretization errors become large.

The conclusion that one must use an optimal choice of $m_c$
to allow one to work in the GSM regime
is in agreement with that obtained 
previously using \tmchpt\ in Refs.~\cite{AB04,ShWu2}.
Thus there is no dispute over 
how to proceed practically, and indeed
all recent simulations use an optimal choice for 
$m_c$~\cite{Lewis,Jansen05,nobend05}. 
My observations here
concern the interpretation of the apparent IR divergences,
their implications for the validity of the Symanzik expansion
at small quark masses,
and the theoretical status of calculations using \tmchpt.

\subsection{Non-optimal critical mass}

I consider first a critical mass having an error of $O(a)$,
since in this case \tmchpt\ at LO provides simple examples
of the fate of the IR divergences. What one learns in this case
can then be generalized to the situation with an optimal $m_c$.
One superficial complication when comparing the results of
Ref.~\cite{FMPR} to those of \tmchpt\ is that the former are
given in the physical basis while the latter are in the twisted
basis. Thus I begin by briefly summarizing the approach and results
of Ref.~\cite{FMPR} using the language of the twisted basis.
Their first step is to determine the form of the Symanzik local effective
Lagrangian corresponding to the lattice theory under study.
The target continuum theory is given by the operators of dimension 4,
\begin{equation}
\ell_4 = \bar\psi D\!\!\!\! /\ \psi + \mu  \bar\psi i\gamma_5 \tau_3\psi
\,.
\end{equation}
where my use of the twisted basis shows up in the form of the mass term.
Note that $\mu$ is the physical quark mass (called $m_q$ in Ref.~\cite{FMPR}).
The dominant discretization errors come from the dimension 5 contribution,
which, in the twisted basis, is
\begin{equation}
a\ell_5 \sim  a\bar\psi i \sigma_{\mu\nu} F_{\mu\nu} \psi +  a\LQ^2 \bar\psi \psi
            +  a\mu^2 \bar\psi\psi
\,.
\label{eq:L5res}
\end{equation}
Here the notation is schematic, showing only the form and order of
magnitude of each term, but omitting the unknown coefficient of $O(1)$
which multiplies each term, and which can have a logarithmic dependence on $a$.
The final term in $\ell_5$, being proportional to $\mu^2$, does not contribute
to the leading apparent IR divergences, and can be dropped from the
subsequent discussion. 

At this stage the analysis can be seen to be
identical to that used in the \tmchpt\ approach, as described,
for example, in Ref.~\cite{ShWu1}.
There is, however, one subtlety in this comparison.
The result in eq.~(\ref{eq:L5res}) apparently differs
from that of Ref.~\cite{ShWu1} by
the absence of the $a\LQ^2\bar\psi \psi$ term in the latter.
This can be traced to the definition of critical mass used in Ref.~\cite{ShWu1}:
the quantity denoted $\tilde m_c$ is defined to include all perturbative
{\em and non-perturbative} contributions to the untwisted mass term.
In other words, in the approach of Ref.~\cite{ShWu1},
the $a\LQ^2\bar\psi \psi$ term in $\ell_5$ is moved into $\ell_4$ and
absorbed by an $O(a)$ shift in the definition of the untwisted quark mass.
Conversely, maximal twist in the definition used by Ref.~\cite{FMPR} corresponds
to having an untwisted mass $m\sim a\LQ^2$ in the $\ell_4$ of
Ref.~\cite{ShWu1}, and this is naturally moved to $\ell_5$.
In fact, in Ref.~\cite{ShWu1} there is an additional term in $\ell_5$
of the form $a m^2\bar\psi\psi$, but this is now seen to be
of size $a^3\LQ^2\bar\psi\psi$, and thus resides in $\ell_7$.
An additional operator in Ref.~\cite{ShWu1} of the form 
$m \bar\psi D\!\!\!\! /\ \psi\to \LQ \bar\psi D\!\!\!\! /\ \psi$
can be absorbed by a redefinition of the fermion field.

The \tmchpt\ analysis and that of Ref.~\cite{FMPR} now diverge.
In the former it is noted that the operators in $\ell_5$ transform
under $SU(2)$ chiral symmetries like mass terms, and thus it is known from
the methods of chiral perturbation theory how to systematically
incorporate their effects
into the chiral Lagrangian describing the low-energy physics of QCD.
At LO, the transcription is
\begin{equation}
a\ell_5 \longrightarrow -\hat{a} \frac{f^2}{4} \Tr(\Sigma+\Sigma^\dagger)
\,,
\label{eq:l5tochpt}
\end{equation}
where, as above, $\hat{a}=2 W_0 a$, with $W_0\sim\LQ^3$ an unknown LEC.
The r.h.s. of eq.~(\ref{eq:l5tochpt})
has exactly the same form as an untwisted mass term, with a mass of size $a\LQ^2$,
in the chiral Lagrangian.\footnote{%
At this point the choice of whether the $a\LQ^2\bar\psi\psi$ term was placed
in $\ell_4$ or $\ell_5$ becomes irrelevant, since the total coefficient
of $\Tr(\Sigma + \Sigma^\dagger)$ is the same in both cases.
}
Thus if one studies the theory as a function
of $\mu$, one is working along a line of the type denoted ``method (iv)''
in Fig.~\ref{fig:maxtwist}.
In \tmchpt, the effects of $\ell_5$ can be completely determined at LO
by simply incorporating into the target continuum theory of $\ell_4$ an
untwisted mass $\delta m=\hat{a}/(2B_0)$. 
The resulting theory has no IR divergences
(since the total quark mass $m_q=\sqrt{\mu^2 +(\delta m)^2}$ does not
vanish) and one can obtain expressions for physical quantities
valid for all $\mu$. In particular, the remaining contributions from
$\ell_5$ [i.e. those of NLO in \tmchpt\ for which the transcription
into the effective chiral theory is more complicated than that
in eq.~(\ref{eq:l5tochpt})], as well as those from $\ell_6$ and higher
order terms in the Symanzik expansion, lead to corrections suppressed
by the expected powers of $a\LQ$~\cite{AB04,ShWu2}. This holds true
also for the properties of particles other than pions~\cite{WW}.
%Thus the Symanzik expansion remains valid, with the caveat that
%the leading order term, $\ell_5$, must be treated non-perturbatively.

I return now to the analysis of Ref.~\cite{FMPR}.
This is based on the
fact that $\ell_5$ (and $\ell_7$, etc.) creates
a neutral pion from the vacuum (recall that $\bar\psi\psi$ in the twisted
basis corresponds to $i\bar\psi_{\rm phys}\gamma_5\tau_3\psi_{\rm phys}$
in the physical basis), as well as giving rise to vertices among odd numbers
of such pions, and to matrix elements in which other hadrons couple to odd
numbers of neutral pions. The resulting neutral pion propagators at
zero four-momentum give factors of $1/m_\pi^2$. A detailed accounting
of these contributions leads to the conclusions given above concerning the
presence and form of IR divergences. In particular, if $m_\pi^2 \sim a$,
then all the leading IR divergences proportional to $(a/m_\pi^2)^{2k}$
are of the same size, suggesting a breakdown in the Symanzik expansion.
Note, however, that these leading IR divergences result only from multiple
insertions of $\ell_5$, and do not involve higher order contributions
to the Symanzik effective Lagrangian ($\ell_6$, $\ell_7$ etc.).
This is an indication that a summation of all apparent IR divergences
may be possible.

Indeed, my observation here is that \tmchpt\ provides such a summation.
This is accomplished, as outlined above, simply by absorbing the additional
$O(a)$ term into the original mass term in the chiral Lagrangian.
Note that since $\delta m\sim a$ one is necessarily in the GSM regime.
The shift in the mass causes a change in the vacuum: at LO, the
twist angle of the condensate becomes $\omega_m=\tan^{-1}(\mu/\delta m)$
rather than $\pi/2$. The apparent IR divergences are then seen as 
an indication that one is expanding around the wrong vacuum. Of course,
it would be possible to force the vacuum in \tmchpt\ to remain at
$\omega_m=\pi/2$, and treat the term in eq.~(\ref{eq:l5tochpt}) as
a perturbation. This would reproduce the diagrammatic analysis
of Ref.~\cite{FMPR}.
%, for the latter is precisely the development
%of an effective theory of neutral pions (the charged pions also present
%in \tmchpt\ not playing a role in the summation). 
In any case, the
main point I want to make is that the summation provided by \tmchpt\
shows that there is no breakdown of the Symanzik expansion, 
in the sense that it is legitimate to treat subsequent terms
as having progressively smaller contributions.
The subtlety here is that one must treat $\ell_5$ non-perturbatively,
so that the manifest powers of $a$ associated with $\ell_5$ are lost.
But once this is done, as is possible with \tmchpt, the manifest
powers of $a$ from subsequent terms ($\ell_6$ etc.) are retained.

Let me illustrate these general observations with two concrete examples.
It is shown in Ref.~\cite{FMPR} that the most divergent quantities are the
overall correlation functions, in contrast to particle energies, where the
divergence is ameliorated by having one less factor of $1/m_\pi^2$.
Thus I consider the two-point function of the ``physical'' axial current,
$A^1_{\mu,{\rm phys}}$, at large Euclidean separation, 
so that the charged pion contribution dominates. 
In the approach of Ref.~\cite{FMPR}, one proceeds as though one is
at maximal twist, and thus the physical axial current is,
in the twisted basis, given purely by the vector current:
$A^1_{\mu,{\rm phys}}= V^2_\mu$. Thus the correlation function of interest is
\begin{equation}
\langle A^1_{\mu,{\rm phys}}(x) A^1_{\mu,{\rm phys}}(0)\rangle
\equiv \langle V^2_\mu(x) V^2_\mu(0) \rangle
\,.
\label{eq:AAtoVV}
\end{equation}
Here, for simplicity, I am assuming that both lattice axial and vector currents
are multiplied by appropriate $Z$-factors so that they are correctly normalized.
Using the results of Ref.~\cite{ShWu2}, the $VV$ correlator can be evaluated
in \tmchpt. It is sufficient for illustrative purposes to work to LO. Then one
finds
\begin{equation}
\langle V^2_\mu(x) V^2_\mu(0) \rangle = 
\sin^2\omega_m
 \langle \widehat{A}^1_\mu(x) \widehat{A}^1_\mu(0) \rangle
+ \textrm{non-pion contributions}
\,,
\end{equation}
where $\widehat{A}$ is the actual LO physical axial current appropriate
to a vacuum oriented such that $\tan\omega_m=\mu/\delta m$. The overall
factor in front of the correlator enters directly in the decay constant
(squared), and so one finds, at LO:
\begin{equation}
\frac{f_A(\textrm{method (iv)})}{f} = \sin\omega_m 
=  \frac{\mu}{\sqrt{\mu^2+(\delta m)^2}} 
=1 - \frac12\left(\frac{\delta m}{\mu}\right)^2 %+ O\left(\frac{\delta m}{\mu}\right)^4
+ \dots
\,.
\label{eq:fAres}
\end{equation}
Since, at LO, $m_\pi^2 \propto \mu$, the corrections to
unity have precisely the IR divergent forms deduced by Ref.~\cite{FMPR}.
As can be seen, however, they sum up into a simple,
predictable form.\footnote{%
At NLO in the GSM regime, the result from method (iv) can be shown
to give $\sin\omega_m$ times the NLO result obtained in Ref.~\cite{ShWu2}.
It is this NLO form which is plotted below in Fig.~\protect\ref{fig:IR}.}

The second example is the charged pion mass. At LO in \tmchpt\ the
result is simple:
\begin{equation}
m_{\pi^\pm}^2(\textrm{method (iv)}) = 2 B_0 \sqrt{\mu^2 + (\delta m)^2}
= \hat\mu \left(1 + \frac12\frac{(\delta m)^2}{\mu^2} + \dots\right)
\,.
\label{eq:mpires}
\end{equation}
Thus the corrections to the squared charged pion mass begin at
order $(\delta m)^2/\mu$, one power of $1/\mu$ less infrared divergent
than the corrections to $f_A$. This (as well as the
form of the higher order terms) is as
predicted by Ref.~\cite{FMPR}. Once again, however, the apparently
IR divergent terms are summed up straightforwardly.

Before turning to the issue of apparent IR divergences when
using an optimal $m'_c$, I think it useful to show
examples of the expected behavior of the quantities just discussed
for various choices of $\delta m$. To do so I use the full
NLO results, valid in both GSM and Aoki regimes,
rather than the LO expressions for the GSM regime given in
eqs.~(\ref{eq:fAres}) and (\ref{eq:mpires}). 
This gives a more realistic view of the expected forms.
Figures~\ref{fig:IR} and \ref{fig:IRaoki} show
the results of \tmchpt\ for $f_A/f$ and $m_{\pi^\pm}^2$.
The parameters are as for Figs.~\ref{fig:mpi}
and \ref{fig:PCAC} ($\delta_W=\delta_{\widetilde W}=-0.3$,
$2w'=(0.25\;\textrm{GeV})^2$, $w_3=0$, $L_i=0$ and chiral logarithms
dropped), except that I have added a non-vanishing
continuum analytic part to $f_A$ (setting $L_4+L_5/2=0.007$)
to make the slope versus quark mass more realistic,
and that the sign of $w'$ is changed for Fig.~\ref{fig:IRaoki} so as to
show the behavior for the Aoki-phase scenario. 
For my choice of parameters the end-points in the first-order scenario 
(Fig.~\ref{fig:IR}) lie at $(m'',\mu)\approx (4,\pm 12.5)\;$MeV 
while the Aoki-phase (Fig.~\ref{fig:IRaoki}) runs from
$m''=-5\;$MeV to $+20\;$MeV. This substantial asymmetry is due to
the relatively large size of $\delta_W$.

\begin{figure}
\centering
\subfigure[$\,f_A/f$]{
\label{fig:IR:fa}
\scalebox{1}[1]{\includegraphics[width=3in]{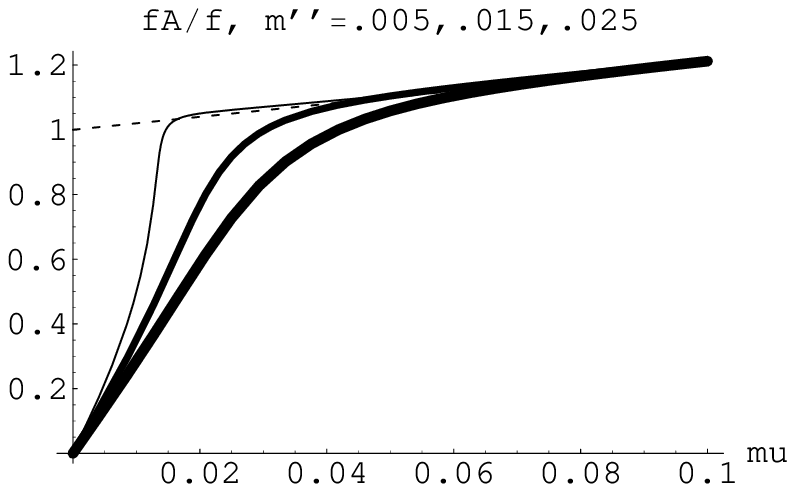}}}
\hspace{0.5in}
\subfigure[$\,m_{\pi^\pm}^2$]{
\label{fig:IR:mpi}
\scalebox{1}[1]{\includegraphics[width=3in]{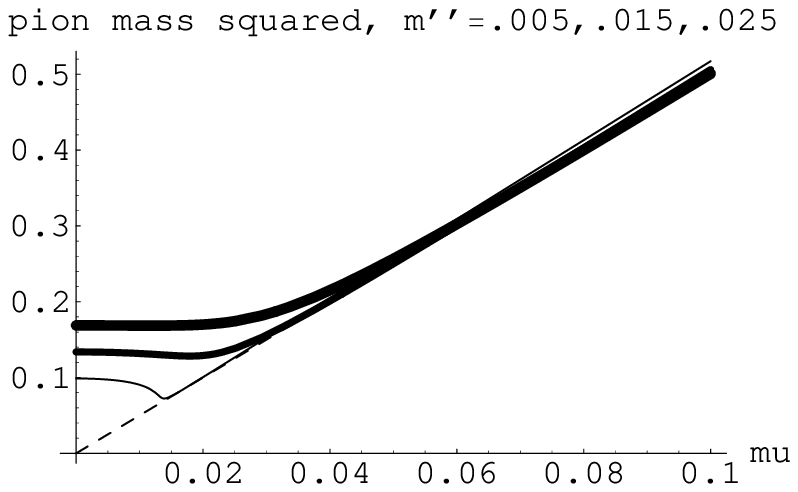}}}
\caption{\label{fig:IR}
Examples of quark-mass dependence when using a critical
mass with an error of $O(a)$
[method (iv) of Fig.~\protect\ref{fig:maxtwist}]: 
(a) $f_A/f$ and (b) $m_{\pi^\pm}^2$ (in GeV${}^2$) versus $\mu$ (in GeV) 
The parameter-set is described in the text, and
corresponds to the first-order transition scenario.
$m''$ is fixed to $\delta m$, with values  $0.005$,
$0.015$ and $0.025$ GeV. These are distinguished by the line width,
which increases with the magnitude of $\delta m$.
The dashed line shows the continuum result.}
\end{figure}

\begin{figure}
\centering
\subfigure[$\,f_A/f$]{
\label{fig:IRaoki:fa}
\scalebox{1}[1]{\includegraphics[width=3in]{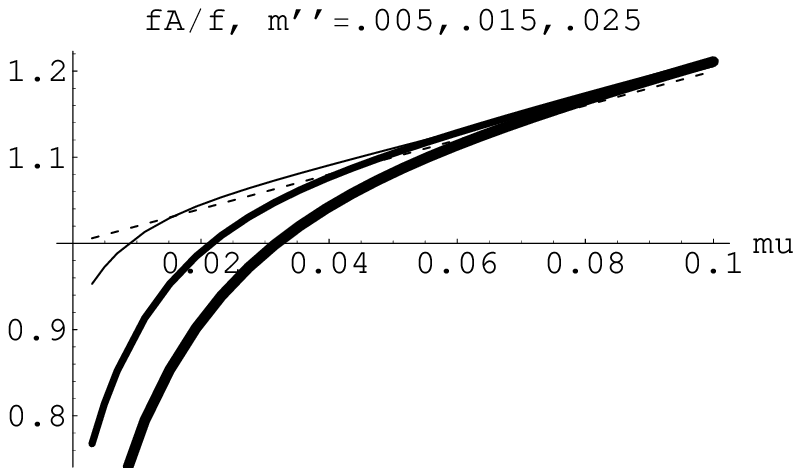}}}
\hspace{0.5in}
\subfigure[$\,m_{\pi^\pm}^2$]{
\label{fig:IRaoki:mpi}
\scalebox{1}[1]{\includegraphics[width=3in]{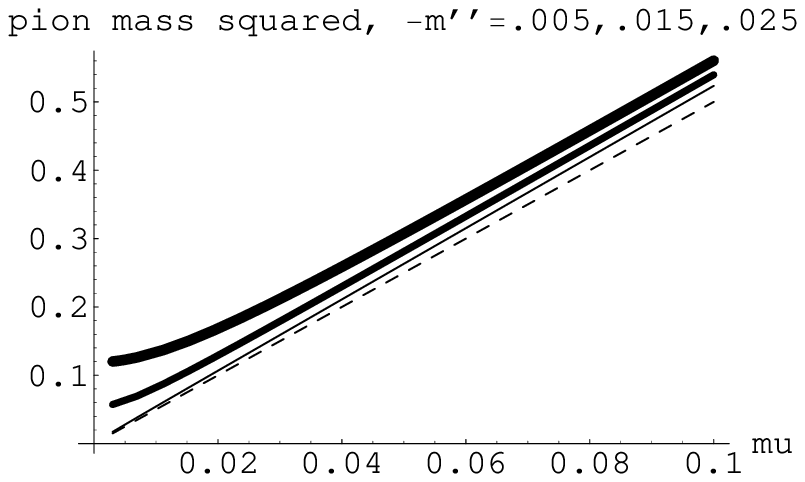}}}
\caption{\label{fig:IRaoki}
As in Fig.~\protect\ref{fig:IR} except that the sign of $w'$ is changed
so that one is in the Aoki-phase scenario. In addition, the values
of $\delta m$ are chosen to be negative in (b).}
\end{figure}

In all the plots the dashed lines are the continuum results
for the parameters I have chosen, assuming that one is at
maximal twist. The solid lines illustrate
the dependence on the twisted mass $\mu$ for three choices of
the untwisted mass $\delta m$. 
I use $\delta m=25$, $15$ and $5\;$MeV, 
except in Fig.~\ref{fig:IRaoki:mpi}, 
where I use $-25$, $-15$ and $-5\;$MeV.
These values are chosen to bracket the GSM and Aoki regimes
(recall that $a^2\LQ^3=7.5\;$MeV for $a^{-1}=2\;$GeV).
The largest value of $|\delta m|$ (giving the curves
with the thickest lines) is most representative
of method (iv), i.e. it has $\delta m\sim a$ for typical choices of $a$.
The curves for $|\delta m|=25\;$MeV bend away from the
continuum form for $\mu \ltapprox \delta m$, as expected
from eqs.~(\ref{eq:fAres}) and (\ref{eq:mpires}).
The NLO corrections are significant, however, as can be seen
from the fact that the results are different for
first-order and Aoki-phase scenarios.
In other words, the results depend on the choice of $w'$, 
which is a NLO parameter in the GSM regime, 
rather than just on $|\delta m|$ and $\mu$, as predicted
by the LO results.

I include the results for $|\delta m|=5\;$MeV to
illustrate what happens if one enters into the Aoki regime. 
These do not correspond to method (iv) as defined above,
and the expressions (\ref{eq:fAres}) and (\ref{eq:mpires}) 
are not valid when $\mu \approx |\delta m|$.
This can be seen clearly from the $\delta m=5\;$MeV curves
for the first-order scenario in Fig.~\ref{fig:IR}:
they are not smooth, and there is a
very large difference between them and those for
the Aoki-phase scenario. 
I choose negative values for $\delta m$ in Fig.~\ref{fig:IRaoki}
so that only the $\delta m=-5\;$MeV result corresponds to running
into the Aoki phase---indeed this curve shows what happens when
one runs into an end-point of the Aoki phase, while those
for $\delta m=-15$ and $-25\;$MeV avoid the phase boundary.
The deviations from the continuum result are also larger 
for negative than for positive $\delta m$ (another indication
that the LO expressions are insufficient).

The intermediate choice $|\delta m|=15\;$MeV lies at, or close to,
the boundary of the Aoki regime, depending on the value of $a$,
and I include it for completeness.

Although the discussion in this section is mainly theoretical,
it is interesting to know whether method (iv) has been used
in practice, i.e. whether the error in $m_c$ is of $O(a)$
or smaller. This is not entirely clear.
The traditional method for determining $m_c$ has been to
extrapolate $m_\pi^2$ to zero along the Wilson axis with
a simple fit function starting from relatively large quark masses.
There are two issues that arise in evaluating this ``$m_\pi=0$'' method.
First, even if one did a perfect extrapolation, the result
would differ by $O(a^2)$ from an optimal critical mass~\cite{ShSi,AB04}.
Second, the long extrapolation introduces a systematic error which,
while not parametrically of $O(a)$, may be numerically of $O(a)$
in a given simulation. For example, in the case where there is
an Aoki phase (as appears to be true for quenched simulations with
the Wilson gauge action), the extrapolation might miss the actual
end-point by an amount of $O(a)$. If so, one would be using
what I am calling method (iv). 

The quenched studies of Refs.~\cite{Biet04,Lewis,Jansen05,nobend05}
compare the results at maximal twist using the ``$m_\pi=0$'' definition 
of the critical mass to those obtained with an optimal choice
(using methods (i) or (ii), or, in the case of Ref.~\cite{Biet04},
results from overlap fermions). The results using the former
definition are found to bend away from those with an optimal $m_c$
for $\mu\ltapprox a$.
In light of the previous discussion, however, it is unclear whether
the ``$m_\pi=0$'' definition corresponds to method (iv), and thus
one does not know which of the curves in Figs.~\ref{fig:IR} 
or \ref{fig:IRaoki} best illustrate the expected behavior.
For example, in the $\beta=6$ data of Ref.~\cite{Lewis}, the
value of $\delta m$ is found to be approximately
$0.01 (Z_m/a)\approx 20\;$MeV. Given that $a\LQ^2\approx45\;$MeV and
$a^2\LQ^3\approx7\;$MeV for this simulation, it is unclear whether
$\delta m$ should be treated as of $O(a)$ or $O(a^2)$.
From a practical viewpoint, however, this is not important.
One should attempt to fit the results to the NLO \tmchpt\ forms
keeping $\delta m$ as a free parameter.

\subsection{Optimal critical mass}

I now return to the case of an optimal $m_c$.
The discussion follows a similar line to that for a non-optimal
$m_c$, but in this case there are no simple analytic forms
to illustrate the summation.

In the analysis of Ref.~\cite{FMPR},
the apparent IR divergences are suppressed because, by
construction, the single pion matrix elements of $a\ell_5$
are reduced from $\sim a$ to $\sim a \mu, a^3$
(and are thus the same size as the matrix element of
$a^3\ell_7$ if $\mu\sim a^2$).
The dominant terms are of the form $a^2 (a^2/m_\pi^2)^k$,
with $k\ge1$, and
so are all equally important when $m_\pi^2\sim a^2$, at which
point they are all $\sim a^2$. These terms apparently
preclude one from working in the Aoki regime, for although
the errors are parametrically small (proportional to $a^2$)
they have a complicated and non-uniform 
dependence on the lattice spacing
and quark mass. Furthermore, the Symanzik expansion is
apparently breaking down.

The \tmchpt\ analysis shows, however, that, as above,
the IR divergences indicate expansion about the wrong vacuum
and can be summed by implementing a non-perturbative
shift to the correct vacuum.
As noted in Ref.~\cite{FMPR}, the leading IR divergences
result from multiple insertions of $\ell_6$,
with only one creation of a neutral pion by
the suppressed vertex $\ell_{5,7}$. Correspondingly, 
in the \tmchpt\ analysis the
change of vacuum is that caused by the $a^2$ coefficient $w'$.
This is exactly the analysis of the phase structure
in the Aoki regime carried out previously
in Refs.~\cite{Mun04,Scor04,AB04,ShWu1}, and extended here
to NLO. The results for $f_A/f$ and $m_\pi^2$ are exemplified
by the $|\delta m|=5\;$MeV curves in 
Figs.~\ref{fig:IR} and~\ref{fig:IRaoki}.

The conclusions I draw are as follows. First, there
is no obstacle, in general, to simulating in the Aoki regime.
The \tmchpt\ analysis provides functional forms that
allow one to predict and fit the behavior of physical
quantities in this regime. There is, in other words, no breakdown
of theoretical understanding in this region, and in particular
no breakdown of the Symanzik expansion, in the sense that
once one includes the leading effects of $\ell_6$  non-perturbatively,
the subleading contributions from these operators, and the contributions
of higher order operators (including now $\ell_5$), 
are systematically suppressed by the expected powers of $a\LQ$.
In fact, if one works at maximal twist, defined using any
of methods (i-iii) (with the caveat that one
must stay above the end-point in the first-order scenario),
then the contributions of $\ell_6$ remain perturbative also,
along with their manifest powers of $a$. This is because, when
using any of these methods, the
condensate points in direction of maximal twist 
up to small corrections, $\omega_m=\pi/2 + O(a)$.

Second, although the apparent IR divergences are summed up
by \tmchpt, their ``residue'' is
a complicated dependence of physical quantities on $a$ and $\mu$.
Third, and most important from a practical point of view, if
one works at maximal twist using any of methods (i-iii),
then the continuum extrapolations are {\em not complicated}.
Indeed, as already noted in the previous section, there are no $a^2$
(or $a^3$)
corrections to the charged pion masses in this case.
Thus in the Aoki-phase scenario, there is no obstacle to working
all the way down to $\mu=0$, although in the first-order scenario
one must stop at $|\mu|\gtapprox 2w'$. Nevertheless, the overall point is
that one does not need to impose the constraint 
$m_q\sim\mu > a^2$~\cite{AB04,ShWu2}.

There is an exception to the claim that \tmchpt\ controls the
IR divergences. This is in the vicinity of the end-points of
the phase-boundary. Here, as discussed in the previous section,
the power counting of \tmchpt\ breaks down,
and higher order terms, including loops, are needed to determine
the vacuum. Thus, in the first order scenario, one should work
at maximal twist only until one is a distance  $\delta\mu\sim a^3$
away from the transition.

\subsection{Subleading IR divergences}

The analysis of Ref.~\cite{FMPR} points out that, in addition
to the leading order IR divergences, there are subleading IR
divergent contributions. Although the former have been understood
and summed using \tmchpt, what of the latter? I have not done
a complete analysis of this question, but I think the above analysis
of the leading divergences makes the following conjecture reasonable:
the subleading IR divergences are removed by a small shift in the
vacuum, perturbatively calculable in \tmchpt\ (except near
the end-points). This incorporates
the NLO and higher order effects of $\ell_5$ and $\ell_6$,
as well as the contributions of higher order operators. 
 In other words, I conjecture
that there is no breakdown of the Symanzik expansion, 
in the sense described above, as long as
one uses \tmchpt\ to include the dominant terms ($\ell_5$ in the
GSM regime and $\ell_6$ in the Aoki regime) non-perturbatively.
Progressively higher order terms ($\ell_7$, $\ell_8$, etc.) 
will then have progressively smaller effects, each subsequent order
suppressed by $a\LQ$.

This brings me to the logical relation between the approach 
of Ref.~\cite{FMPR} and \tmchpt. I want
to reiterate that the starting point of \tmchpt\ is the
Symanzik expansion, but that \tmchpt\ goes beyond by including
our non-perturbative understanding of spontaneous chiral symmetry breaking.
Thus it is based on a tower of two local effective theories: 
the first describing
the long-distance physics of lattice QCD, and
the second describing the long-distance behavior of the first.
The theoretical status of the second effective theory, i.e. chiral
perturbation theory, is, perhaps, less solid than the first, since one
cannot do a perturbative analysis following Symanzik~\cite{Symanzik}
as chiral symmetry breaking is a non-perturbative phenomenon. 
Nevertheless, the theoretical basis for \tmchpt, which is identical
to that for continuum $\chi$PT, is strong~\cite{Weinberg,Leutwyler}. 
The main shortcoming of \tmchpt, I believe, is not its
theoretical foundation, but rather the practical need to truncate
the expansion at NLO or, perhaps, NNLO, and the related issue of
how well the expansion converges for present quark masses and lattice
spacings.
In this regard, the present analysis and the conjecture  made above
are important, because they imply that there is no non-uniformity
in a joint expansion in $m_q$ and $a$. The absence of IR divergent
terms implies that one makes only small errors by truncating the
Symanzik expansion at, say, $\ell_7$ (as is done in the previous section).

\section{\label{sec:conc} Conclusions}

Studies of tmLQCD are still at a relatively early stage,
and it is important to fully understand the practical and
theoretical issues that are involved. Key issues are the
determination of the critical mass and the size, and chiral
behavior, of the remaining discretization errors. Particularly
important discretization errors are those giving rise to
parity and flavor breaking, for unless these are small in practice
they will likely lead to complications in extracting
phenomenologically important hadronic quantities.

Here I have focused on the theoretical issues, showing how
by applying chiral perturbation theory to Symanzik's effective
Lagrangian one can systematically study the properties of
discretization errors. The aim here is to understand the errors
so that they can be removed, or reduced, in a controlled way.
For example, if one wants to work at maximal twist, where the
errors of $O(a)$ are automatically removed, then it is useful
to know what happens when the choice of critical mass has errors
of a given size. The forms given in Ref.~\cite{ShWu2} and 
generalized here provide such information to NLO in the joint
chiral-continuum expansion.
In particular, I have extended previous results in the Aoki regime
to NLO, allowing consistent NLO fits for all quark masses.

On a practical level, there is general agreement
in the literature on how to proceed:
one should use one of the ``optimal'' non-perturbative definitions
of maximal twist [methods (i), (ii) or (iii)]. 
Here I have clarified the relation between the different definitions,
both in the GSM and Aoki regimes.

There is however, disagreement on the interpretation of 
what happens to discretization errors as one approaches the
chiral limit.
With  non-optimal choices of the critical mass 
(including the ``$m_\pi=0$'' definition 
irrespective of the accuracy of the extrapolation)
the results for a number of physical quantities
exhibit a ``bending" effect, in which the results diverge
away from the expected chiral behavior for $m_q \ltapprox a$.
These can be understood qualitatively in \tmchpt\ as a result
of working at a non-maximal twist angle, with the divergence
from $\pi/2$ increasing toward the chiral limit~\cite{ShWu2}. 
Examples of the expected forms are given in Figs.~\ref{fig:IR} and
\ref{fig:IRaoki}. A seemingly different interpretation has
been given in Ref.~\cite{FMPR}, in which the bending is
due to discretization errors which are proportional
to inverse powers of the quark mass, and which, for $m_q\sim a$,
arise from contributions proportional to
all powers of $a$. I have shown, however, that the interpretations
are consistent: the apparently IR divergent errors of Ref.~\cite{FMPR}
are simply an expansion of the convergent expressions of \tmchpt,
and indicate the need to do a non-perturbative
shift in the vacuum. 

A similar discussion applies to the discretization errors
that arise when one works with an optimal choice of maximal
twist. The IR divergent errors analyzed in Ref.~\cite{FMPR} are
summed up by \tmchpt\ applied in the Aoki regime.
One consequence is that there is no barrier to working at maximal
twist in this regime, i.e. with $m_q \sim a^2$, as long as one 
uses the optimal choice of maximal twist determined in this regime
(rather than by extrapolation from larger masses) and as long as
one does not work all the way down to the second-order end-points 
if one is in the scenario with a first-order transition.

It is suggested in Ref.~\cite{FMPR} that the apparent IR divergences
signal a breakdown of the Symanzik expansion at small quark masses
($m_q\sim a^2$ for an optimal choice of maximal twist).
I have argued that this is not the case, although the situation
is somewhat subtle. 
The apparent IR divergences are an indication that, in general,
the leading
terms in the Symanzik expansion (those proportional to $a$ and $a^2$) must
be treated non-perturbatively, and this is what is accomplished
by \tmchpt. This might be interpreted as a breakdown in the
simple expansion in powers of lattice spacing, since insertions of
any number of factors of $\ell_5$ and $\ell_6$ are required. On the other hand,
one can treat subsequent terms, $\ell_7$, $\ell_8$, etc., as
small corrections, so in this sense the expansion remains intact.
In fact, at maximal twist, defined optimally, the contributions of
$\ell_5$ and $\ell_6$ can also be treated perturbatively, so the
Symanzik expansion remains valid in the usual sense that all powers of
$a$ are manifest. It is for this reason that automatic improvement
at maximal twist remains valid into the Aoki region (and all the
way down to $m_q=0$ in the Aoki-phase scenario~\cite{AB04,ShWu2}).
This result should simplify numerical applications of tmLQCD.

After this paper was completed, Ref.~\cite{AB05} appeared, containing
results overlapping those in secs.~\ref{sec:maxtwist},
\ref{sec:fits} and \ref{sec:aoki}, as well as detailed
fits of numerical results to \tmchpt\ expressions.

\section*{Acknowledgments}
I am very grateful to Oliver B\"ar, 
Roberto Frezzotti, Maarten Golterman, Giancarlo Rossi and
Jackson Wu for detailed
comments and discussions.
This research was supported in part by 
U.S. Department of Energy Grant No. DE-FG02-96ER40956.

\end{document}